\documentclass[journal=langd5,manuscript=article]{achemso}

 \usepackage{float}
 \usepackage[labelfont=bf,labelsep=period]{caption}
 \usepackage[english]{babel}
 \usepackage{sectsty}
 \usepackage[utf8]{inputenc}
 \usepackage{array}
 \usepackage{geometry}
 \usepackage{setspace}
 \usepackage{xkeyval}
 \usepackage{xcolor}
 \usepackage{natbib}
 \usepackage{textcomp}
\usepackage{comment}
 
 \usepackage{siunitx}
 
\author{Xuehua Zhang}
\affiliation{Department of Chemical and Materials Engineering, University of Alberta, Alberta T6G 1H9, Canada}
\email{xuehua.zhang@ualberta.ca}

\author{Zixiang Wei}
\affiliation{Department of Chemical and Materials Engineering, University of Alberta, Alberta T6G 1H9, Canada}

\author{Howon Choi}
\affiliation{Department of Chemical and Materials Engineering, University of Alberta, Alberta T6G 1H9, Canada}

\author{Hao Hao}
\affiliation{Central Faculty Office(FSET), Swinburne University, Melbourne 3122, Australia}

\author{Huaiyu Yang}
\affiliation{Department of Chemical Engineering, Loughborough University, LE11, 3TU, UK}
\email{H.Yang3@lboro.ac.uk}

\title {Oiling-out Crystallization of Beta-Alanine on Solid Surfaces Controlled by Solvent Exchange}

\begin{document}
 
\begin{abstract}
\footnotetext{Z.W. and H.C. are equal contributors.}

Droplet formation in oiling-out crystallization has important implication for separation and purification of pharmaceutical active ingredients by using an antisolvent. In this work, we report the crystallization processes of  oiling-out droplets on surfaces during solvent exchange.  Our model ternary solution is beta-alanine dissolved in isopropanol and water mixture. As the antisolvent isopropanol displaced the alanine solution pre-filled in a microchamber, liquid-liquid phase separation occurred at the mixing front.  The alanine-rich subphase formed surface microdroplets that subsequently crystallized with progression of solvent exchange. We find that the flow rates have significant influence on the droplet size, crystallization process, and growth rate, and final morphology of the crystals. At fast flow rates the droplets solidified rapidly and formed spherical-cap structures resembling the shape of droplets, in contrast to crystal microdomains or thin films formed at slow flow rates. On a highly hydrophilic surface, the crystals formed thin film without droplets formed on the surface. 
 We further demonstrated that by the solvent exchange crystals can be formed by using a stock solution with a very low concentration of the precursor, and the as-prepared crystals can be used as seeds to trigger crystallization in bulk solution. Our results suggest that the solvent exchange has the potential to be an effective approach for controlling oiling-out crystallization, which can be applied in wide areas, such as separation and purification of many food, medical, and therapeutic ingredients.

\end{abstract}

\section{Introduction}

Crystallization is one of the most important separation and purification techniques used in the pharmaceuticals and chemical engineering to produce a wide range of pure materials\cite{myerson2002handbook,kleinebudde2017continuous}. A method in crystallization from a liquid mixture is antisolvent crystallization, where the crystal nucleation and growth is driven by solubility reduction with change in the solvent composition  \cite{li_2016_process, yu_2005_effects, nowee_2008_antisolvent}. Antisolvent crystallization is usually at low cost and with high efficiency.  In a range of crystallization systems, oiling-out or liquid-liquid (L-L) phase separation frequently occurs, which is the appearance of droplets in supersaturated solution. Many hydrophobic pharmaceutical compounds exhibit oiling-out behavior \cite{kiesow_2010_solubility}, such as active pharmaceutical ingredients (butyl parben, antiandrogen, erythromycin ethylsuccinate and idebenone), food compounds (vanillin and lauric acid), and some polymer such as polyethylenglycoldimethylether \cite{lafferrre_2004_study, suzuki_2019_a, bhamidi_2019_a, yang2014ternary, yang2012investigation, ilevbare_2013_liquidliquid}.  Recent development of many hydrophobic pharmaceutical compounds including new anticancer and antivirus drugs has led to increasing interest in controlling oiling-out crystallization \cite{serajudin_2002_salt, dealbuquerque_2017_influence, sun2020understanding, tanaka2019effect, wang2020emulsion}.

Droplet formation from oiling-out may be used for  obtaining different crystal shapes \cite{bonnett_2003_solution,sun_2018_oilingout, sun_2018_understanding}, producing spherical crystals \cite{pitt_2018_particle, kawashima_1982_spherical}, or separation of fatty acids from the organic phase \cite{maeda_1997_separation}.  The oiling-out  crystallsiation can induce different morphologies of crystals, including spherical particles \cite{pitt_2018_particle, sun2019design, duffy2012situ} and porous particles \cite{yang2014sandwich}.  On the other hand, oiling-out crystallization is often viewed as undesirable because droplets may degrade the quality crystals from higher impurities and reduce the yield of the products due to the loss of the desired compounds to the droplets \cite{deneau_2005_an}. The droplet formation increases the complexity of the crystallization process, as there are phase changes from liquid-liquid (L-L) phase to liquid-liquid-solid (S-L-L) phase and from liquid-liquid-solid phase (L-L-S) to liquid-solid (L-S) phase \cite{kiesow_2010_solubility, mcclements_2012_crystals, yang2012investigation}. 
L-L phase region in the ternary phase diagram is mainly dependent on the temperature and the composition of the mixture.  For given pair of the good solvent and the antisolvent, the solvent ratio is essential to control the path of the phase separation. Solubility phase diagram of the compound, the good solvent and the the antisolvent provides a guideline for determination of the solvent ratio for crystallization \cite{yang2014ternary, sun_2018_oilingout, codan2010phase, yang2014sandwich, goetsch2016superposition}. Lately, researchers started looking closely into the interface between the two liquid phases \cite{tatsukawa2020development, ianiro2019liquid, walton2019using}, where the mass transfer and heat transfer occur, directly influencing formation of the droplets and crystallization.

Up to now, what rarely explored is to understand and control oiling-out crystallization under flow conditions by solvent exchange. Advantages of the solvent exchange process have been demonstrated clearly in controlled formation of multicomponent nanodroplets on surfaces \cite{zhang2007,zhang_2015_formation,bao2016,li2019, li2019small}. 
 The solvent exchange process has been performed to ternary liquid mixtures with Ouzo effect (i.e. spontaneous emulsification induced by solvent dilution) \cite{zhang2007,zhang_2015_formation}. As an oil solution in a good solvent is displaced with an antisolvent,  a transient oversaturation of oil is created from the increasing ratio of the antisolvent at the diffusive mixing front, leading to the nucleation and growth of the droplets. The final droplet volume after the solvent exchange strongly depends on the flow rate of the antisolvent \cite{zhang_2015_formation,bao2016,dyett2018}. For given flow conditions, the nucleation and growth of the droplets can be fine tuned down to sub-femtoliter by adjusting the initial concentration and micropatterns on the surface \cite{bao2015}. As such, the solvent exchange is complementary to other processes for controlling the liquid-liquid phase separation, such as drop evaporation \cite{yang2015phase,li2020}.

The solvent exchange process also provides a way to understanding the transition from oiling-out droplets to crystals. 
In particular, the dynamics of oiling-out crystallization of individual droplets is difficult to access a bulk mixture with L-L phase separation.  As two layers of liquid phase form, with agitation, a huge amount of droplets form as dispersed phase \cite{yang2012investigation, sun2020understanding}, rendering the mixture cloudy where the droplets inevitably interact with each other. During the solvent exchange, the droplets are immobilized on the surfaces, and the dynamics of the droplet formation and growth and of crystallization from L-L phase separation can be monitored in-situ without interference from too many other droplets around.  Furthermore, the crystals formed by the solvent exchange are immobilized on the surfaces, which may be useful as ingredients in surface functionality or as templates for film synthesis. 

In this work, we will investigate oiling-out crystallization during the solvent exchange. Our model system is beta-alanine in a binary solvent mixtures of water and isopropanol. The ternary phase diagram of beta-alanine, isopropanol and water is available from literature \cite{shanthi_2013_nucleation,sun_2018_oilingout}, which assists us to determine the solution composition. Alanine was chosen as the model compound also because it is a simple non-chiral molecule so the complexity in enantiomeric interactions or polymorphism will not need to be considered in this work. We will monitor the formation of droplets on surfaces, and growth rate and evolution of crystals in-situ with time. We will show that the solvent exchange can be an effective approach for controlling oiling out crystallization for obtaining diverse forms of solidification such as thin film and spherical caps by simply adjusting flow rate. 
The general principle presented in this work may be applied to crystallization of many other pharmaceuticals compounds with oiling-out behaviours.

\section{Experimental section and theoretical analysis}

{\bf Chemicals and materials}

The stock chemicals, beta-alanine (ACROS organics, 99 \%), isopropanol (Fisher Scientific, 99.9 \%) were used as received in our experiments. {Ethanol reagent (Fisher Scientific, HPLC Grade, including 90 \% ethanol, 5 \% methanol, and 5 \% isopropanol), was used in cleaning substrates.} The solutions were prepared by mixing water, alanine and isopropanol and sonicating till all alanine powder fully dissolved. The solutions were stored in a sealed container for overnight before use for the experiments.  The composition in the solution was 3\% beta-alanine, 45 \% water, and 52 \% isopropanol by weight and the antisolvent was isopropanol for the solvent exchange. 

Silicon and silicon dioxide substrates were purchased from University Wafer (South Boston, MA, US).  The layer of wet thermal oxide ($SiO_2$) on the surface was 300 $nm$ in thickness. $SiO_2$ wafer was cut manually by a diamond scriber into 1.3 $cm$ by 2 $cm$ then cleaned by sonication in a beaker of deionized water (from Mili-Q Direct) and ethanol for 10-20 minutes each before use. The substrate were dried by compressed air before being placed on the bottom of the flow cell.  

A highly hydrophilic surface was prepared by treating bare silica substrates in the piranha solution for 10-20 minutes at 75 $^oC$ then sonicated in deionized water for 10-20 minutes. (Caution: Piranha solution is highly caustic.) The hydrophobic substrate (OTS Si) were bare silica surface coated with octadecyltrichlorosliane (OTS, Alfa Aesar, 99.9 \%). The coating was prepared by following the procedure reported in previous work \cite{zhang2008}. 

{\bf Solvent exchange for oiling-out crystallization}

 The {schematic} design of the flow cell and the experimental set up can be seen in Figure \ref{digram}(A). { The solvent exchange was performed in a microfluid chamber, presented in our previous work.\cite{fluidchamber}}. The chamber was $\sim$ 350 $\mu$m in height, 1.3 cm in width and 4.5 cm in length. A clear solution consisting of water, isopropanol and alanine filled the flow cell. Then the solution was displaced by the antisolvent isopropanol.  
 The flow rate of the antisolvent was controlled by a syringe pump. The solvent exchange completed 1-2 minutes after all droplets on the surface of the substrate crystallized.{The crystals that formed by solvent exchange on substrate surface can be remove by sonication in water (alanine is high soluble in water).}
 
  An upright optical microscope (Nikon H600l) equipped with a camera was used to record the process of the droplet formation and crystallization during the solvent exchange. The droplet size with time was analyzed by ImageJ. 
  Video analysis was used to track the change
of the area of crystals with solvent exchange. We selected
video frames with a fixed interval based on the rate
in crystal growth. We tracked
these changes and calculated the area value based on image
pixels. {The beta-alanine crystals on substrate surface were collected by knife scraping for characterization and seeding application without milling.} {Samples of the crystals on the surfaces were used to determine the XRD (Rigaku) and IR (Agilent Cary 620).}



{\bf Estimate the supersaturation level based on the growth rate of droplets }

We are not able to measure directly the supersaturation, {$S_{droplet}$}, for droplet formation during the solvent exchange. However, we can estimate the level in the liquid mixture, based on the growth rate of  droplets on the surface. The droplets are assumed to grow in a constant contact angle mode on a homogeneous substrate by diffusion of the components from the external liquid to the droplets  \cite{zhang_2015_mixed}. In brief, 

\begin{equation}
    \frac{dM}{dt}=\pi R D \cdot (c_{\infty}-c_{s}) \cdot f(\theta)
    \label{1}
\end{equation}

The $M$ is the total mass of the droplet, $R$ is the radius of the surface area covered by the droplet on the substrate, $D$ is diffusion constant, \(\theta\) is the contact angle of the droplet on the substrate, and \(c_{\infty}\) and \(c_{s}\) are the concentration of droplet liquid in bulk and the saturation concentration in the droplet. Function \(f(\theta)\) and \(g(\theta)\) are two geometric factors.

 The following Equation \ref{4} correlates the {$S_{droplet}$} level $\Delta c$ (=\(c_{\infty}\) - \(c_{s}\)) to the droplet radius R and the droplet radius growth rate $\frac{dR}{dt}$. From the change in the droplet size with time, we can estimate the {$S_{droplet}$} level $\Delta c$. 

\begin{equation}
    \Delta c = \frac{3 \rho}{D} \cdot \frac{g(\theta)}{f(\theta)} \cdot R(t) \frac{dR}{dt} = k \cdot R(t) \frac{dR}{dt}
    \label{4}
\end{equation}

Here \(k\) is a constant(=\(\frac{3 \rho}{D} \cdot \frac{g(\theta)}{f(\theta)}\)). $\rho$ is the density of droplet liquid.

{\bf Crystallization in bulk mixture triggered by seeds}

A suspension containing crystals seeds in isopropanol was made by the following process: Alanine was collected from the surface after oiling out crystallization. 0.1g of alanine was added into 50 $ml$ isopropanol and the mixture was sonicated for 10 mins till the solution turned to milky. Then 0.1 $ml$ of the milky suspension was taken out and diluted by 4 $ml$ of isopropanol to obtain the seed suspension with the concentration of alanine particles at $\sim$ $5\times 10^{-5} g/ml$. {The collected crystals from the surface were used} to trigger the bulk crystallization, an aqueous solution of alanine was prepared by dissolving 13.2 $g$ of alanine in 17.6 $ml$ of water. Then the solution was separated equally into two vials. 4 $ml$ isopropanol was added into the ternary solution in one vial that was kept still for crystallization to occur. In the second vial, 4 $ml$ of seeding suspension was added into the solution to trigger crystallization.


\section{Results }

\begin{figure} [htp]
	\includegraphics[width=1\columnwidth]{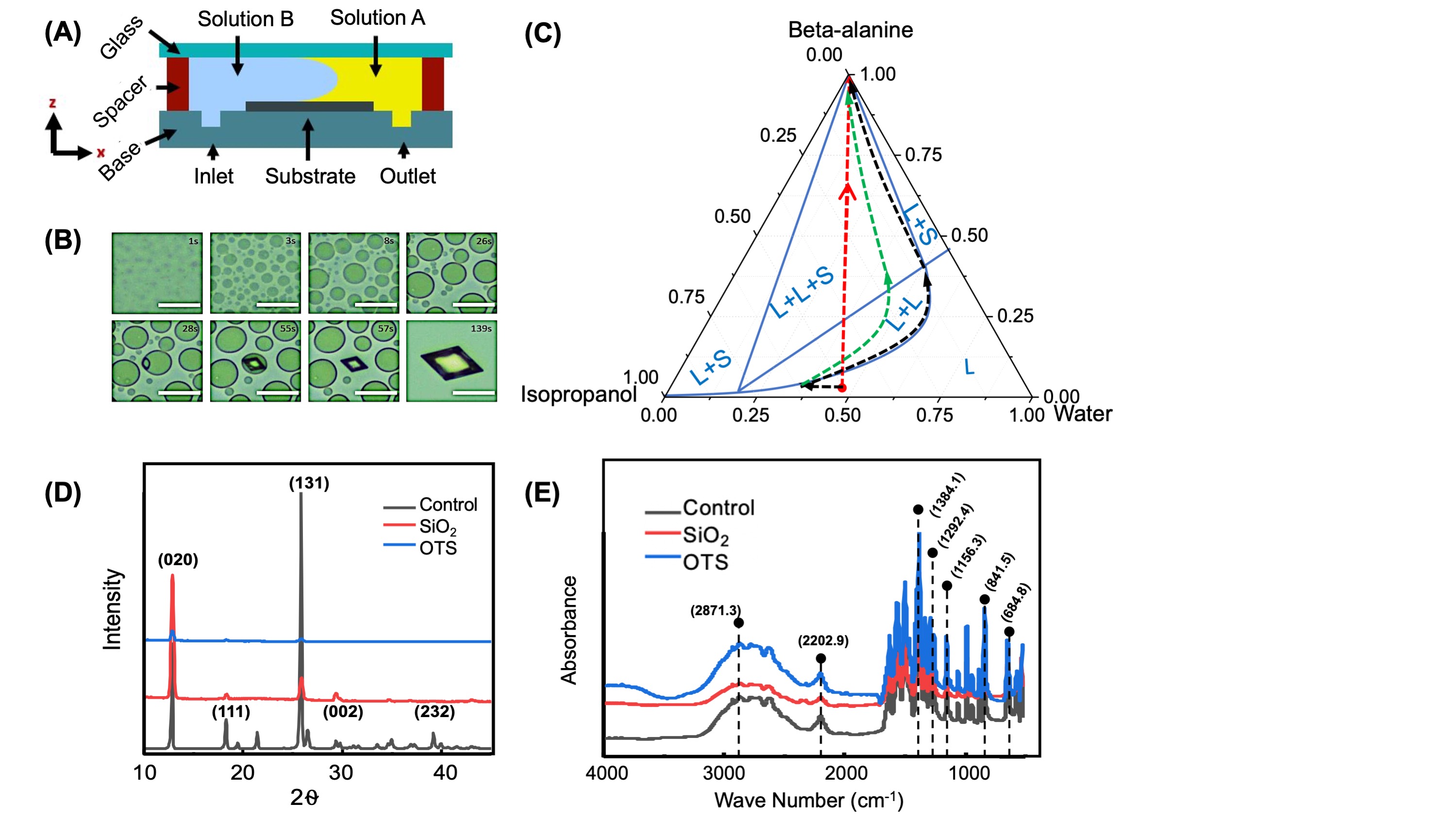}
	\caption{(A) Schematic drawing of the solvent exchange process in a fluid chamber. The antisolvent isopropanol is displacing the solution of alanine in water and isopropanol. (B) Representative optical images of oiling-out crystallization during solvent exchange in case of crystal growth in a single droplet. Length of scale bar: 25 $\mu m$. (C) Schematic pathways of the composition point in ternary phase diagram during the droplet formation and crystallization. Dashed black line: slow flow rate;Dashed green line: fast flow rate.
	(D) FTIR spectrum and (E) XRD of alanine crystals on the substrate and of bulk crystals as control sample. The control sample and thin film XRD of crystals from solvent exchange with the solution of 3 \% beta-alanine, 45 \% water, 52 \% isopropanol and the antisolvent isopropanol with 12 $ml/hr$ flow rate on OTS and $SiO_2$. OTS background is the XRD plot of clean OTS coated silica.}
	\label{digram}
\end{figure}

Figure \ref{digram}(B) represents the process of crystal growth within a droplet. There were two main steps in oiling out crystallization induced by solvent exchange: surface droplet formation from liquid-liquid (L-L) phase separation and crystallization from solid-liquid-liquid (S-L-L) phase separation as shown in Figure \ref{digram}(C). Once formed, the droplets on the surface remained stable in the flow of the antisolvent. {With progression of the solvent exchange in liquid-liquid phase region, the concentration of alanine in the droplets increased till nucleation} and eventually all of the droplets crystallized. The  structure of alanine crystals collected from the surface after the completion of the solvent exchange was characterized by both infrared and X-ray diffraction measurements as shown in Figure \ref{digram}(D)(E). {The same peaks of XRD and IR for all the samples proved that the crystals were same polymorph as bulk alanine crystals. It is noticed that the different intensities of the XRD and IR peaks may be due to different quantities, orientations \cite{paul2020toward}, and crystal sizes \cite{guzzo2019effect} of the measured crystal samples without milling.}

Below we will show that by the solvent exchange various crystal shapes can be created, including thin films, layers, fibers, or microdomes. The size of the droplets from L-L phase separation, and the dynamics in growth and morphology of crystals can be controlled by the flow rate of the antisolvent. The following sections will cover effect of flow rates on silicon dioxide, and then the effects from the substrate properties, and finally a potential application of crystals collected from the surface as seeds for bulk crystallization.

\subsection{Formation of oiling-out droplets}

\begin{figure} [htp]
\includegraphics[width=1.0\columnwidth]{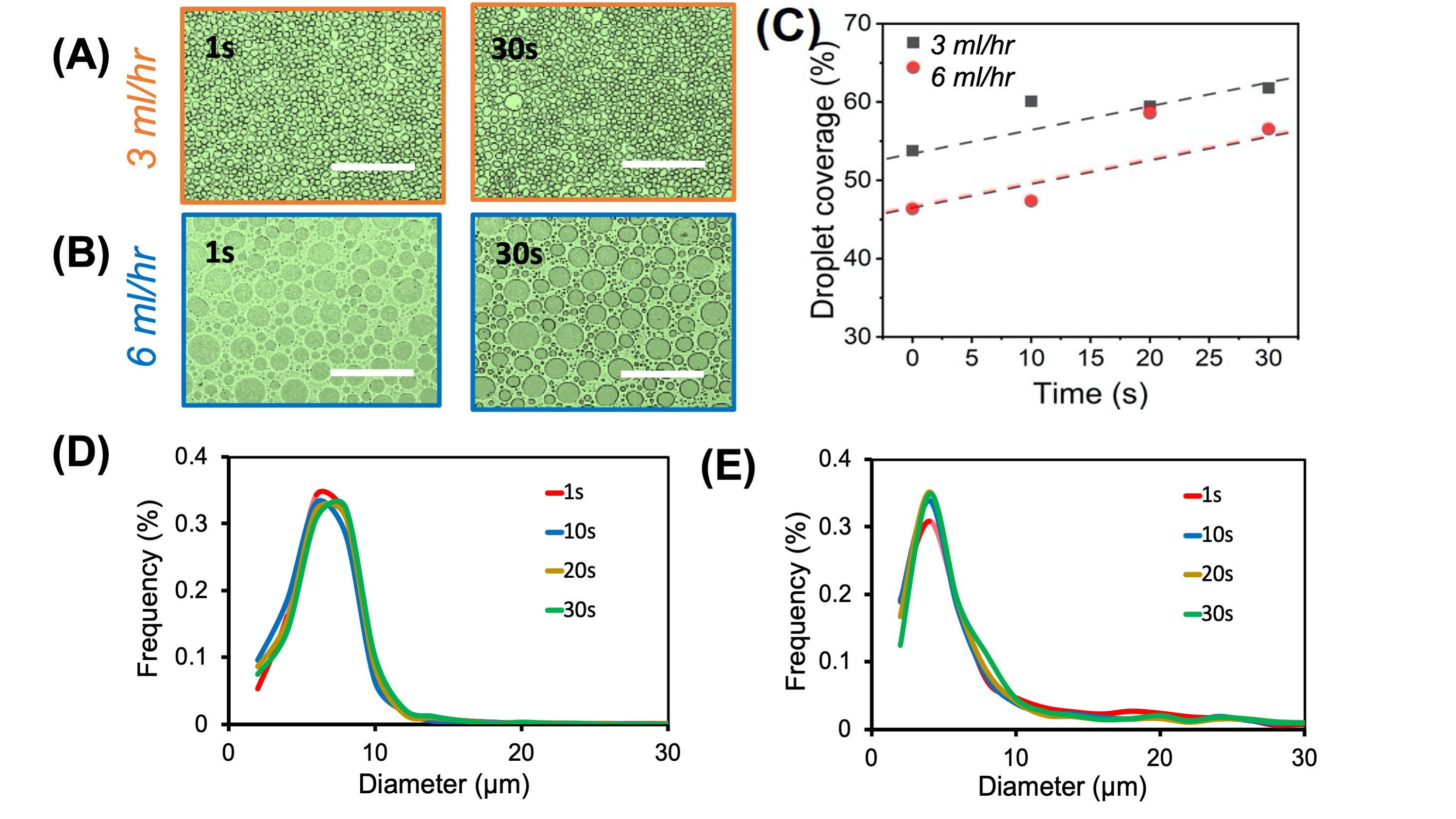}
\caption{Surface nanodroplets within 30 s before formation of first crystal on the surface. The flow rate is (A) 3 $ml/hr$ and (B) 6 $ml/hr$.  Length of the scale bar: 100 $\mu m$. (C) Plot of the coverage area percentage of droplets versus time. Droplets size distribution at flow rate of (D) 3 {$ml/hr$} and (E) 6 $ml/hr$.}
	\label{path}
\end{figure}

\begin{figure}[htp]
\includegraphics[width=1.0\columnwidth]{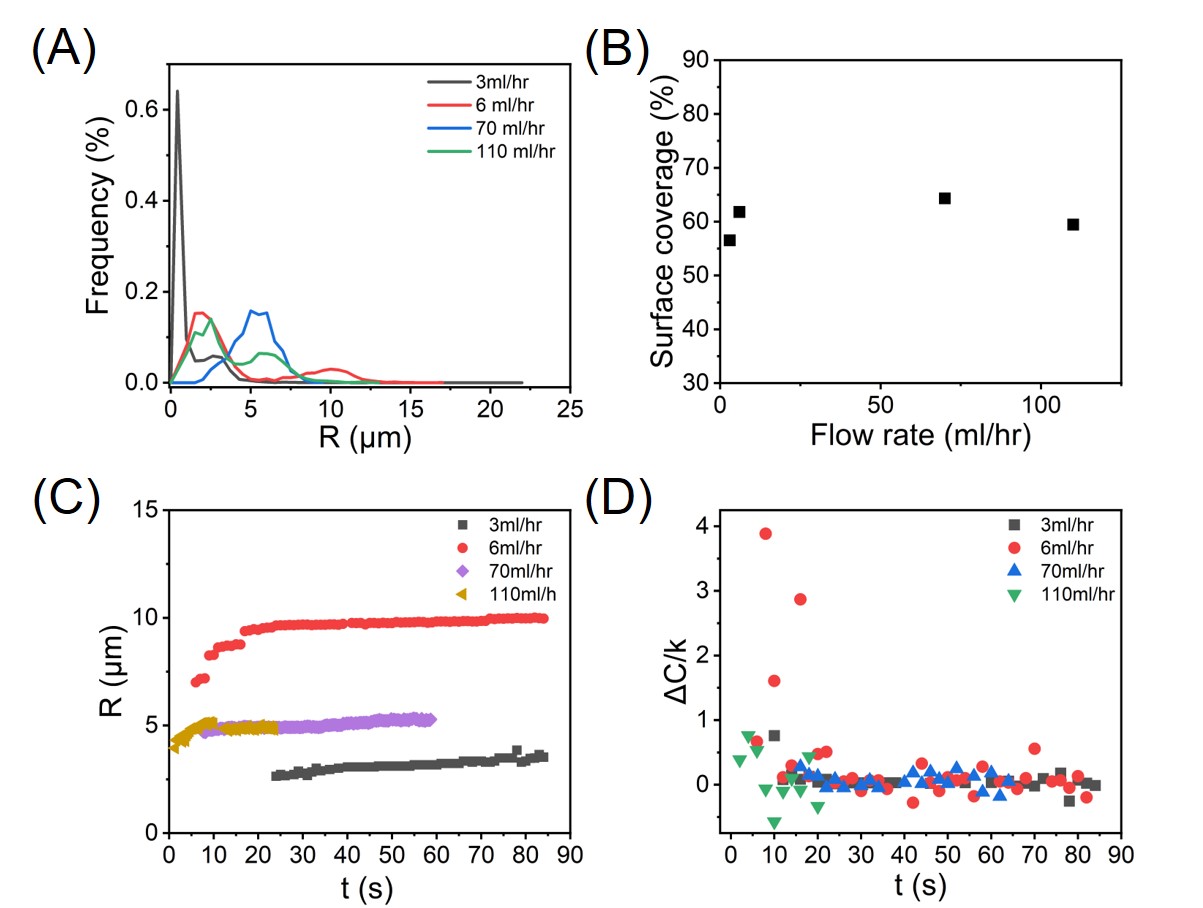}
	\caption{ Plot of the $S_{droplet}$ correlated with time by droplet growth analysis.  (A) Droplet size distribution at different flow rates.(B)The droplets surface coverage at at 3 $ml/hr$, 6 $ml/hr$, 70 $ml/hr$ and 110 $$ml/hr$$. (C)The grow of the droplet with time. (D) $S_{droplet}$ level for the droplet growth with time.}
	\label{oversaturation}
\end{figure}

 The time lapse sequence in Figure \ref{path} shows that at low flow rates from 3 $ml/hr$ to 6 $ml/hr$, the number density and surface coverage of the droplets was high as soon as the droplets was clearly visible in the images (Supporting video 1 and 2). The size distribution of the droplets remained almost constant with time at both flow rates. Bimodal distribution of droplet sizes can be seen at the flow rate of 6 $ml/hr$, while most of droplets were small at 3 $ml/hr$. 
The surface coverage of the droplets reached around 60-70 \% for all flow rates shown in Figure \ref{oversaturation}(B). 
The above results suggest that a large popular of droplets from L-L phase separation mainly took place at the initial stage of the solvent exchange. Absence of continuous L-L phase separation was also evidenced from the transparency of the flow. Otherwise microdroplets would scatter light, giving rise to milky appearance. 

 Based on the size of a single droplet (without coalescence) observed on the substrate, the {$S_{droplet}$} level for the droplet growth can be estimated according to Eq. (1) and (2).  At all flow rates, the radius of an individual droplet is tracked with time, represented in Figure \ref{oversaturation}. Here \(t=0\) is defined to be immediately before the droplet can be detected by the microscope with a spatial resolution of 2$\mu$m. The plot in Figure \ref{oversaturation}(B) shows that the {{$S_{droplet}$}} level was above 0 at the initial 10-20 s, and then remained almost constant with time.  The {$S_{droplet}$} level was estimated based on the growth rate of droplets and plotted in Figure \ref{oversaturation} (C).  
The general trend was that the {$S_{droplet}$} decreases with progression of solvent exchange. Before the crystallization occurred, the supersaturation levels for droplet growth were close to zero at all flow rates.

 The short time window for the droplet growth from oiling out was obviously different from that for oil droplet growth by using an Ouzo solution \cite{lu2016,xue2017}. In the process of solvent exchange by a ternary ouzo solution and an antisolvent, oil droplets formed on the surface, grew and coalescence during the entire period of solvent exchange.  
The continuous growth of the oil droplets was attributed to the {$S_{droplet}$} from the composition in a metastable Ouzo region \cite{lu2016}. For oiling-out droplets, short growth time may be attributed to absence of the metastable region in the solubility phase diagram and the ternary mixture in the droplets.  These oiling-out droplets nucleated and grew from rapid L-L phase separation when the composition is beyond the solubility boundary. With further solvent exchange, the compositions in droplets change, but not the volume of droplets. We will discuss this point further in Discussion section. 
No strong dependence of droplet size with the flow rate may be also explained by the short time for the formation of oiling-out droplets. Such dependence of droplet size on flow rate was different from droplet growth from ouzo solution where the increase in flow rates leads to larger droplets \cite{zhang_2015_formation}. 

It is consistent that  the surface coverage of the droplets was high for all flow rates. Further droplet growth would require the bare surface area released from coalescence of neighbouring droplets that may be delayed as the viscosity of the droplets increase with time. But the delayed coalescence may not be the main contributor for lack of droplet growth, as the short time window is also evident from the clear bulk solution.

\begin{figure}[htp]
			\includegraphics[width=1.0\columnwidth]{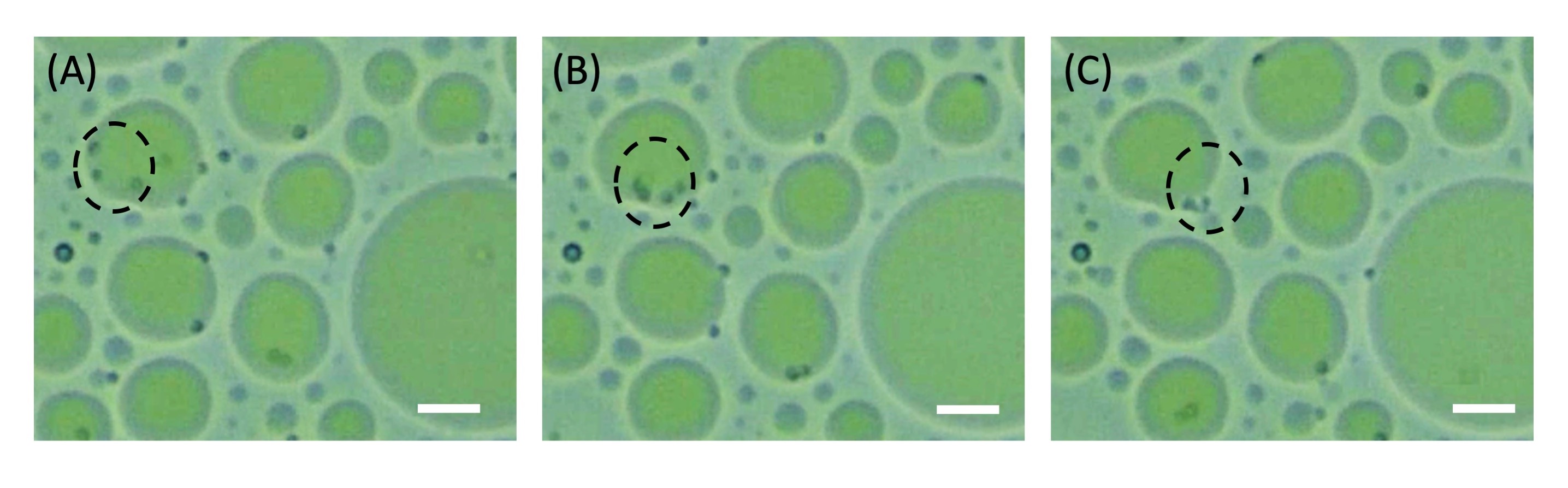}
				\caption{Small droplets of alanine-lean phase (marked by the circle of black dashed line) inside the droplet of alanine-rich phase. The solvent exchange was carried out at flow rate of 70 ml/hr in direction from top to bottom.  Scale bar: 10 $\mu$m. The time period from (A) to (C) is about 3 seconds. }
	\label{pathway}
\end{figure}

The supersaturation level of alanine, {$S_{alanine}$}, inside the droplets increased with progression of solvent exchange, evidenced from formation of small crystal particles inside or around some big droplets. For example, at {6 $ml/hr$} the particles were at the boundary of the droplets as shown in Figure \ref{pathway}. The formation of these crystal particles suggests that the crystallization path of the droplets in the ternary diagram was not along the boundary of L-L phase separation to reach solidification, but inside the region of L-L phase separation to entered S-L-L phase separation as indicated in Figure \ref{digram}(C). However, it is not possible to measure the $S_{alanine}$ inside the droplet and it is difficult to accurately estimate the $S_{alanine}$ based on the ternary phase diagram due to the change of the  compositions for the solution  in the droplets.

\subsection {Crystallization during solvent exchange}

\begin{figure} [htp]
\includegraphics[width=1\columnwidth]{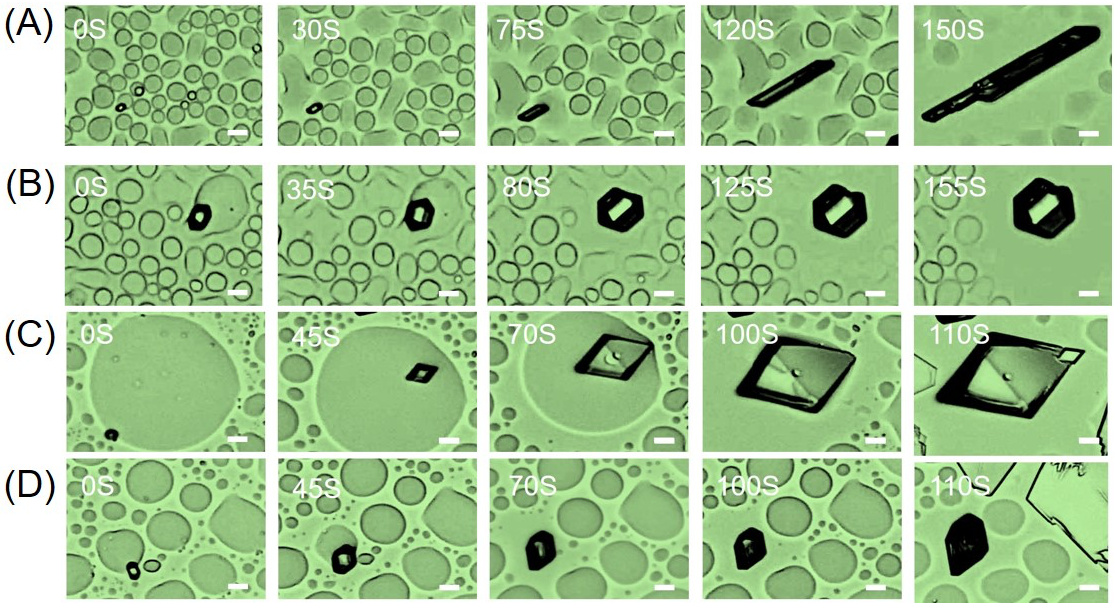}
\includegraphics[width=0.75\columnwidth]{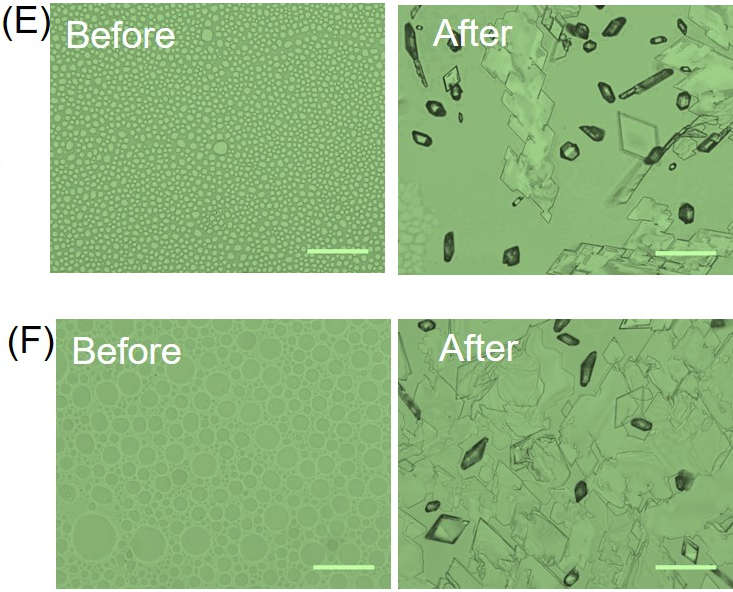}
\caption{Images of crystal growth at 3 $ml/hr$ (A) and (B) and 6 $ml/hr$ (C) and (D) with scale bar of 10 $\mu$m. (E) and (F) are the large view images before and after crystallization at 3 $ml/hr$ and 63 $ml/hr$, respectively, {with scale bar of 100 $\mu$m.} The time of 0 is defined when the first crystal particle appears in the field of view.}
	\label{slow}
\end{figure}

At the very slow flow rate of 3 $ml/hr$, crystallization on the surface can be triggered by some crystal seeds from the flow. These crystal particles came to the field of view and land on the surface sporadically (Figure \ref{slow}).  Afterwards the crystal particles grew very slowly while the droplets around them remain stable. During this period of time, the concentration of alanine inside the droplets may be too low to initiate the crystallization. 
 With further progress of the solvent exchange, the growth of the crystals became more obvious. Landed crystals were observed to grow within the droplets, suggesting that the droplets are rich in alanine. 
 Interestingly, the crystals grew by consuming contents from droplets nearby even without  direct contact.  The droplets around the landed particles shrinked slowly while other droplets remained the same with time, apart from coalescing with each other (Figure \ref{slow}).  
 
 This process was attributed to diffusive interaction between droplets and the growing crystals. {There is an energy barrier for nucleation, while the energy barrier for crystal growth is always considered to be negligible \cite{mullin2012industrial}.} The free energy for crystal growth was much lower than that for crystal nucleation in the droplets.  Therefore, when crystallization occurred in one droplet, alanine from other droplets was transferred into the growing crystals instead of forming new crystals. The mechanism is essentially same as Ostwald ripening,  driven by minimization of free energy. It is only evident at a very slow flow rate when the effect from the external flow can be neglected so that the flux driven by the free energy difference becomes apparent.

  In the regime of slow flow rates, shapes of some crystals resemble those from bulk crystallization.  Three types of common morphology of crystals were observed for the crystals growing from a droplet, namely polygonal shape, needle like shape, and diamond shape (Figure \ref{slow}). Crystals growing in a polygonal shape may be possibly due to the projection of a diamond shape on the plan of 2D images. After consuming all droplet liquid, crystals extended in one direction much faster than in others, triggered crystallization of neighbouring droplets, and formed a long needle like shape. In certain cases, the needle-like crystal may extend at both directions. Crystals also grew in all edges and acquire a regular diamond shape.  The diamond shape was also observed in crystallization from slow evaporation \cite{shanthi_2013_nucleation}. As the growth rate, some crystals can develop regular shapes, while others were influenced by the source of alanine supplied from surface droplets.

\begin{figure} [htp]
 \includegraphics[width=1.0\columnwidth]{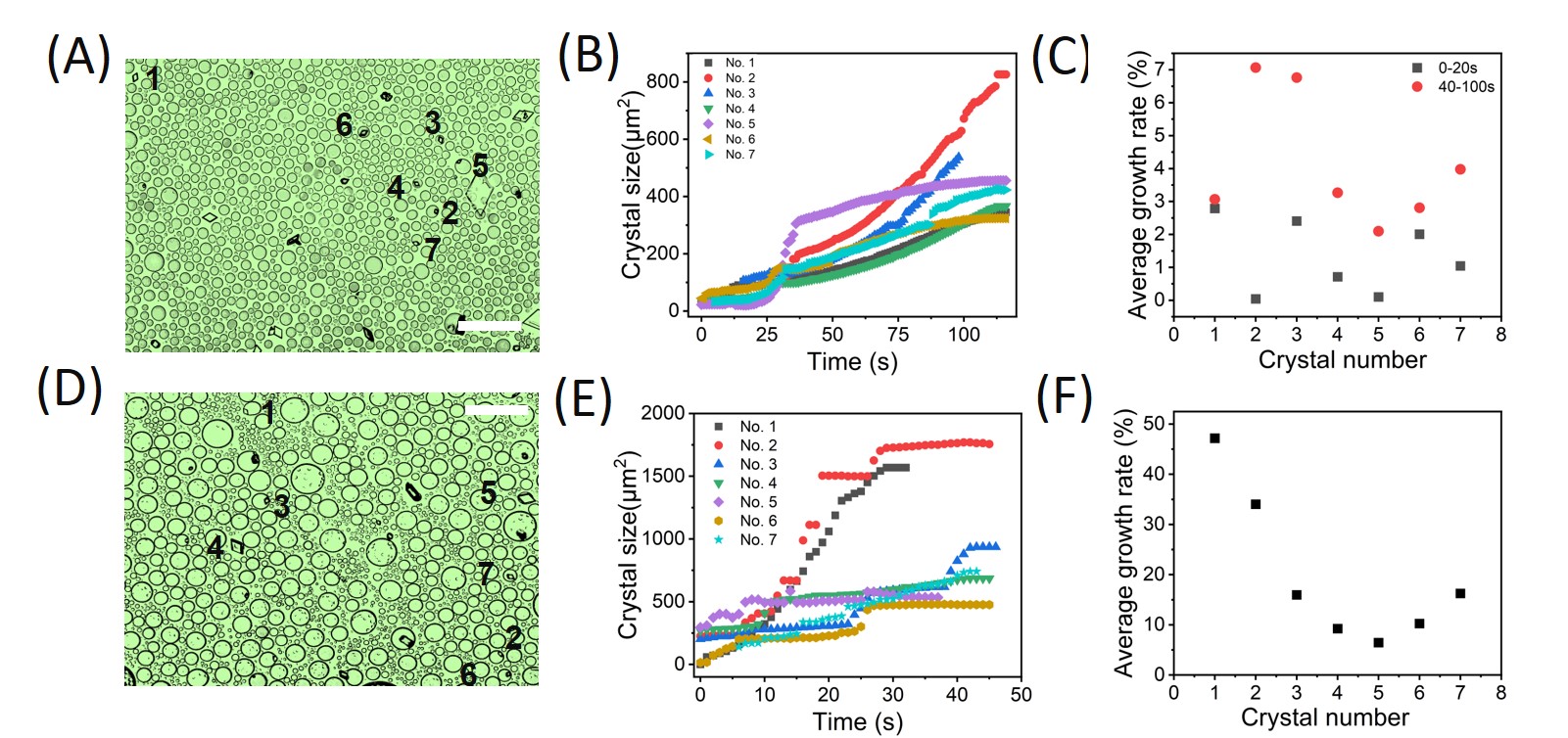}
\caption{(A) Image of the droplet at 5s under flow rate of 6 {$ml/hr$} with scale bar of 100 $\mu m$. (B) Crystal size (coverage area) change of 7 crystal samples on $SiO_2$ substrate during crystal growth period of 100 s under flow rate of 6 $ml/hr$. (C)The average growth rate of those 7 crystals under flow rate of 12 {$ml/hr$}. (D) Image of the droplet at 5s under flow rate of 12 $ml/hr$ with scale bar of 100 $\mu$m.  (E) Crystal size (coverage area) change of 7 crystal samples on $SiO_2$ substrate during crystal growth period of 100 s under flow rate of 12 $ml/hr$.(F)The average growth rate of those 7 crystals at 0-20 s and 40-100 s under flow rate of 12 $ml/hr$.}
	\label{slowrate}
\end{figure}

The early growth rates of crystals were analyzed by tracking the area occupied by the  crystals shown in Figure \ref{slowrate}. Crystals with regularity in their shapes were selected, numbered, and analyzed as representatives in Figure \ref{slowrate}.  The plot shows that at 6 $ml/hr$ crystal growth rate was slow for initial $\sim$ 25 s and was much faster for next 10 s or so. The sharp increase in the growth rate may be related to a high transient {$S_{alanine}$} level by the solvent exchange. The growth rate of crystals was much faster at 6 $ml/hr$ than at 3 $ml/hr$. 
 
\begin{figure} [htp]
\includegraphics[width=1\columnwidth]{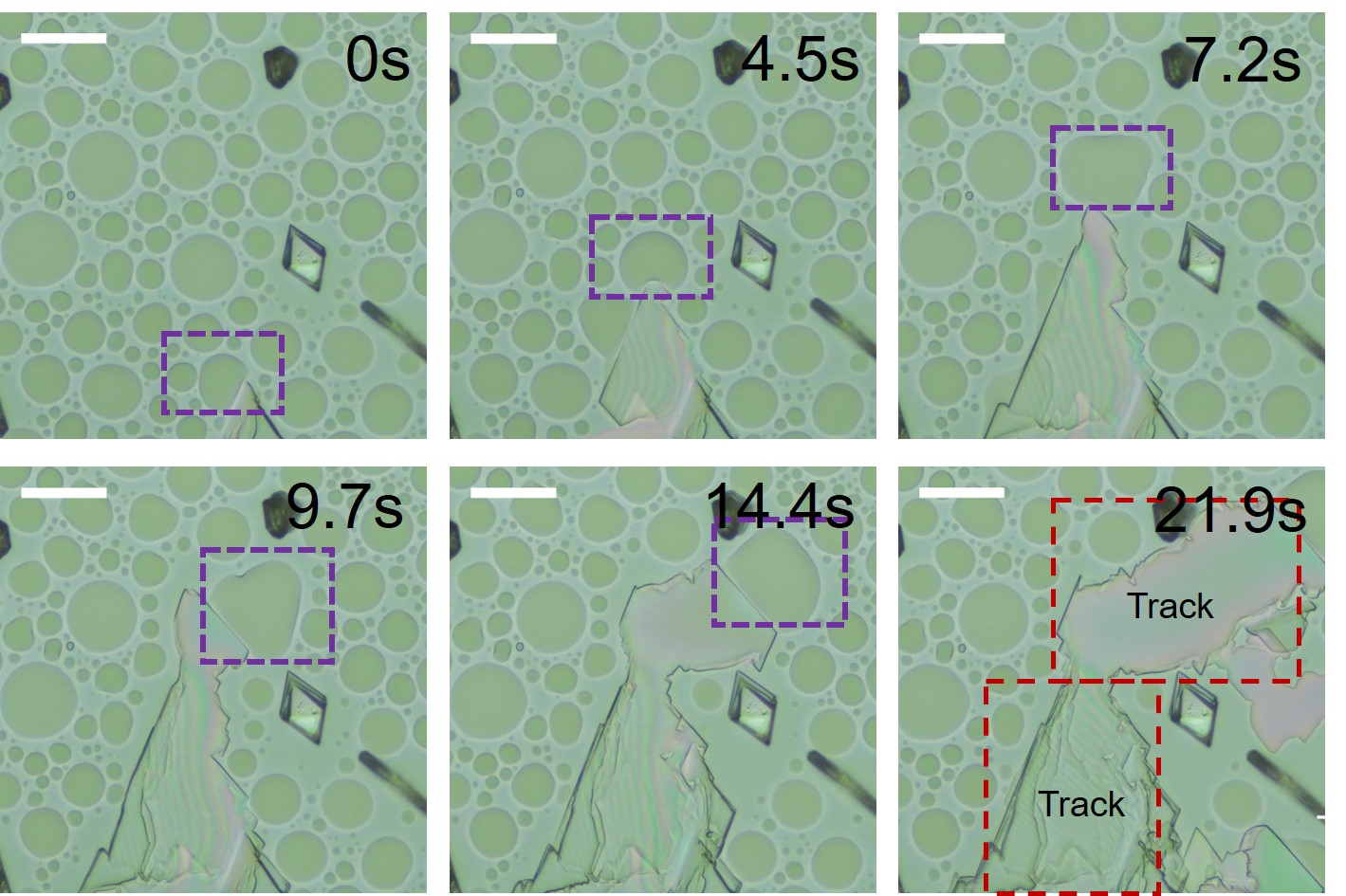}
\caption{Images of self-propelling droplet and the formation of crystal tracks at 6ml/hr. Length of scale bar:50 $\mu$m. Blue dotted box: self-propelling droplet; red dotted box: expanding crystal film. Flow rate: 6 $ml/hr$}
	\label{track}
\end{figure}

 The surface coverage of crystals was dominated by expansion of crystal films at 6 $ml/hr$.  The film of crystals did not follow the position and surface coverage of the droplets. Instead the expanding crystal film pushed the residue liquid of the droplet away from the front while creating a track of the thin crystals on the surface, as shown in Figure \ref{track}.  Such self-propelling droplets was related to the Marangoni effect, that was, the surface tension gradient from local crystallization, similar to crystallization under thermo-mechanical stress \cite{Kant2020}. The droplets moved toward the location of higher surface tension where the liquid had not crystallized. By the self-propelled motion, the area of the crystal films was much higher than the base area of the precursor droplets.

 \begin{figure} [htp]
\includegraphics[width=1\columnwidth]{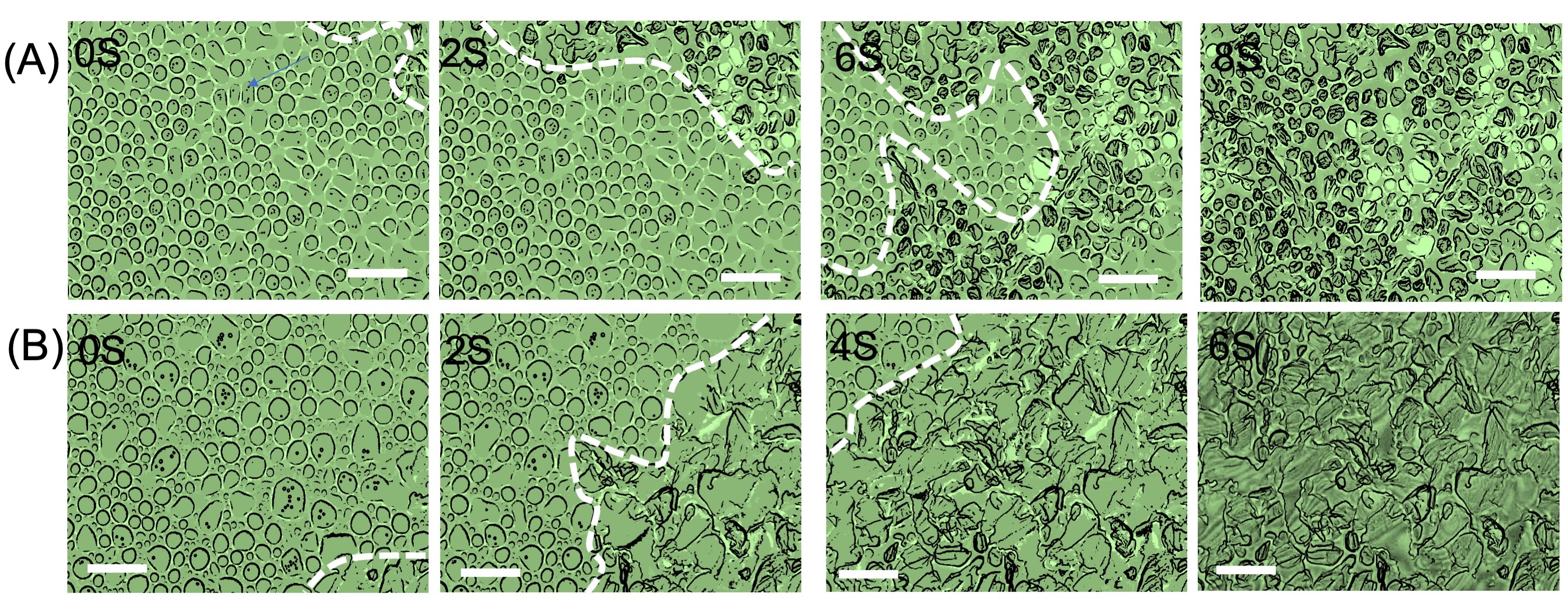}
\caption{Images of crystal growth at 70 $ml/hr$ (A) and 110 $ml/hr$ (B) with scale bar of 50 $\mu m$. For the observation of crystallization, we define \(t_{0}\) to be the initial time of the seed crystal landing on the substrate. }
	\label{fast}
\end{figure}

At flow rate of 70 {$ml/hr$} (Supporting Video 3), the crystallization progressed among individual droplets. In the vicinity of the newly formed crystals, the droplets crystallized without detectable contact with existing crystals as shown in Figure \ref{fast}. Those droplets crystallized at the original location while being compressed towards to liquid side.  The final crystals retained largely the shape and position of the droplets. The crystallization front propagated rapidly among the droplets over the surface. At even faster flow rate of 110 $ml/hr$ (Supporting Video 4), the droplet crystallization was similar to that at {70 $ml/hr$}, except that the crystals were less constrained to the droplet shape. The final structures consisted of crystal film and crystals in the shape of solidified droplets in absence of fibers. {It is noted that
for a certain size (1.3 cm x 2 cm) of the substrate, the crystals in a range of 0.7 - 1 mg, with yield of 20 \% - 25 \%, were formed and collected after solvent exchange.}

 \subsection{Effect of surface wettability}

\begin{figure} [htp]
\includegraphics[width=1\columnwidth]{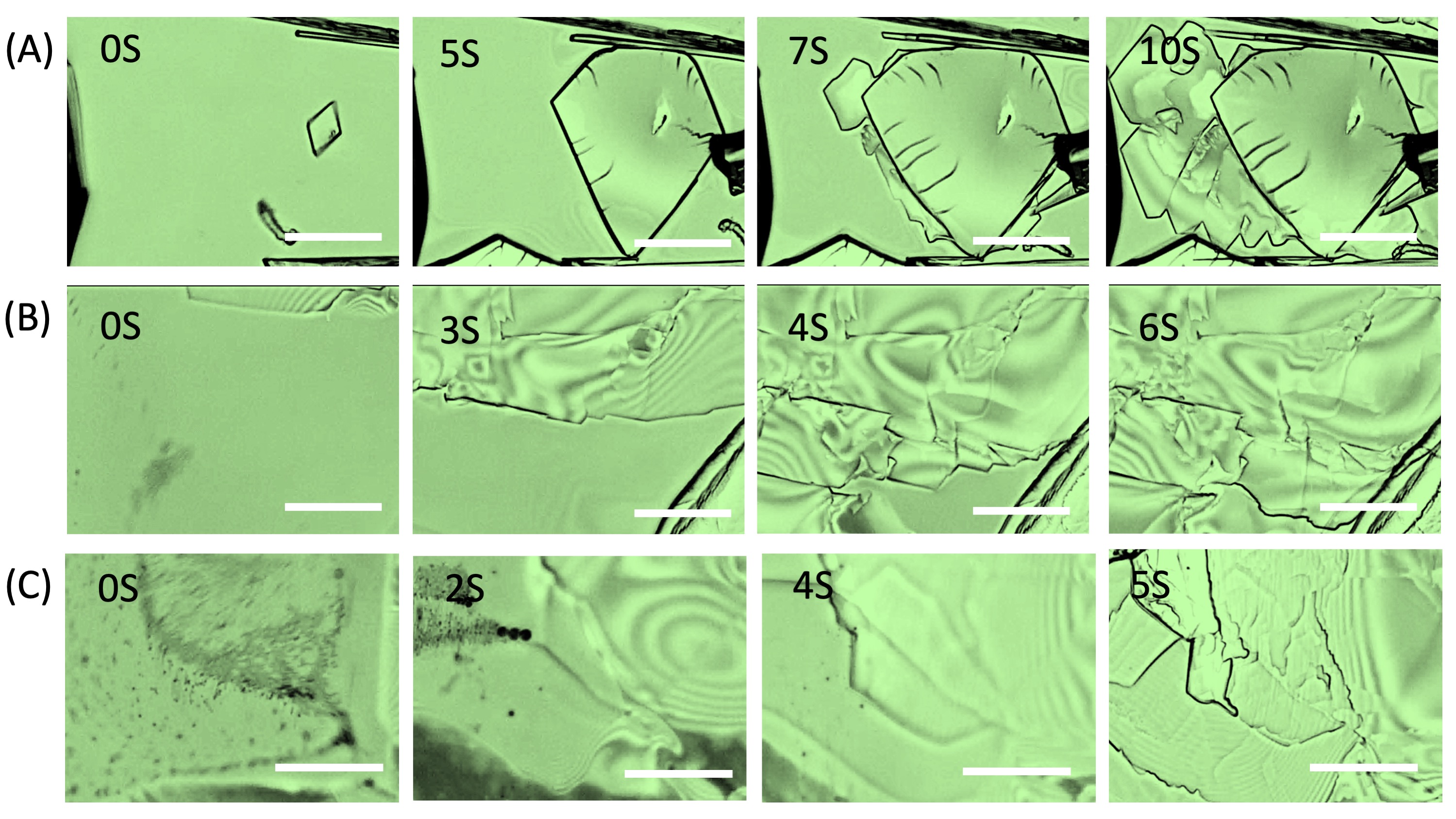}
\caption{Images of crystal growth on Si surface at 12 ml/hr (A) 70 $ml/hr$, (B) and 110 $ml/hr$, (C) with scale bar of 50 $\mu$m. }
	\label{Si}
\end{figure}

On a highly hydrophilic surface of $Si$, the process of crystal formation and growth was clearly different from that on $SiO_2$ described as above. No stable surface nanodroplets formed on the surface before the crystal formation, although many tiny droplets move with the flow suggesting L-L phase separation still took place. Crystal formation initiated randomly on the surface, triggered by a landed crystal or by local perturbation. The area of crystals expanded along the surface in the form of thin film and eventually cover the surface as shown in Figure \ref{Si}. 

\begin{figure} [htp]
	\includegraphics[width=0.95\columnwidth]{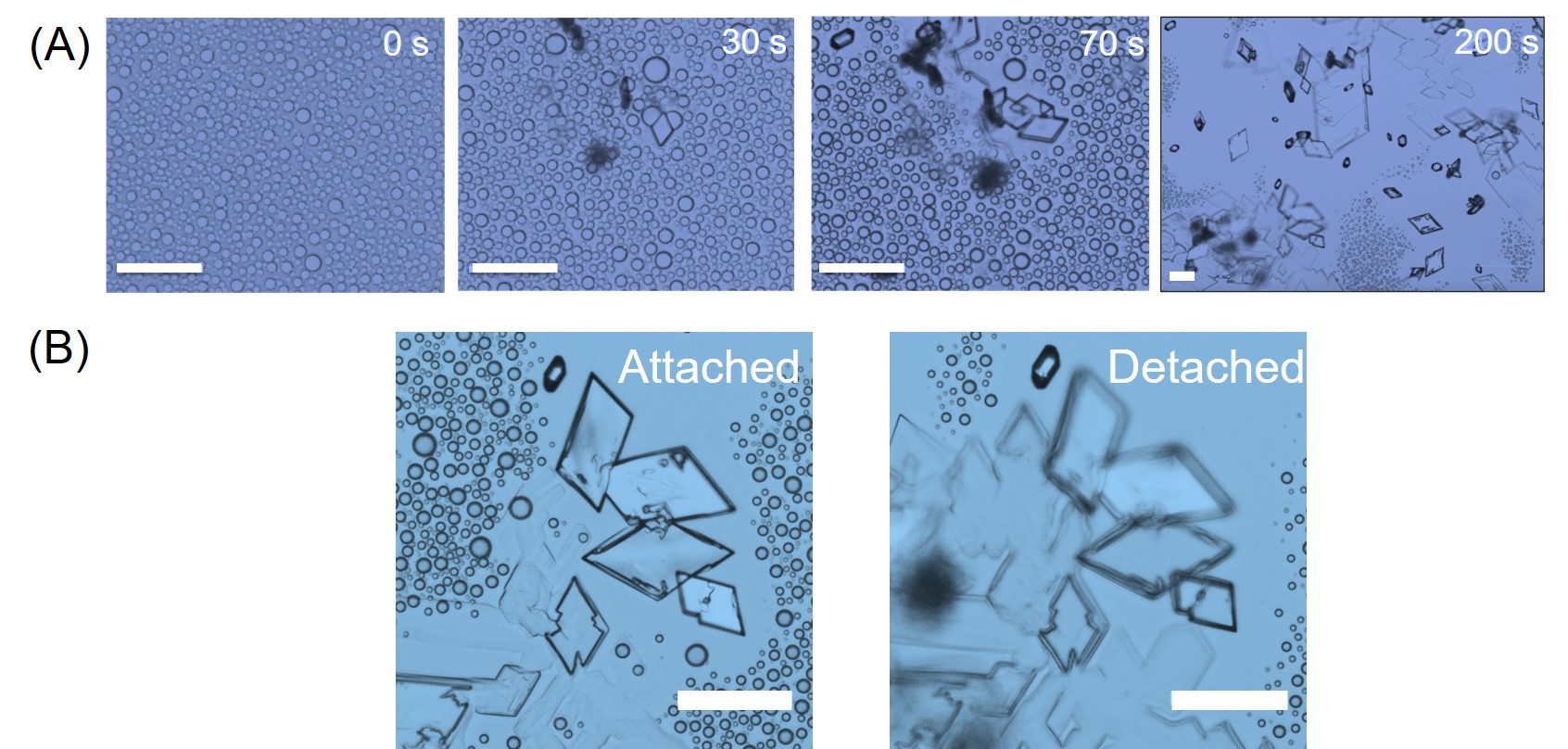}
\caption{(A)Oiling out crystallization on hydrophobic substrate. $t_{0}$ was set to the initial observation of droplets on the substrate. (B)The crystals detached from the hydrophobic substrate surface after crystallization. Flow rate of the solvent exchange:12 {$ml/hr$}. Length of scale bar: 50 $\mu$m.}	
	\label{OTS}
\end{figure}

On the surface of hydrophobic OTS-Si, the behaviour of L-L phase separation and crystallization was similar to that on $SiO_2$. The crystallization was triggered by the crystals land on the substrate and continues to grow. The difference from $SiO_2$ was that the thin film of crystals prefers to spreading in diamond shaped pattern shown in Figure \ref{OTS}. At a higher flow rate, the extension of the film along the surface was faster with less regular diamond shapes. Crystals on $OTS-Si$ can easily {detach} from the surface and moved along with the bulk flow.


\subsection{Discussion}

The solubility phase diagram is essential for understanding multiple phase separations in oiling-out crystallization. As the antisolvent was introduced to displace the alanine solution in the fluid chamber, the ternary composition at the mixing front changed along the dilution path, indicated in the solubility phase diagram in Figure \ref{digram}B. The liquid at the mixing front first did undergoes L-L phase separation, leading to formation of droplets on the solid surface. These droplets consisted of the alanine-rich phase. The alanine-lean phase that formed concurrently with the droplets was transported away by the flow of the antisolvent.


With progression of solvent exchange, the composition in the droplets moved away from the phase boundary between the one-phase liquid region and the two-phase liquid-liquid region as indicated by the dotted line in Figure \ref{digram}(C).  As comparison, the pathway of a ternary droplet from solvent loss by evaporation was indicated by the red dotted line in the phase diagram \cite{sun_2018_oilingout}. When composition point was beyond the boundary between L-L region and L-L-S region, crystallization occurred. Then the composition point moved towards to the point of pure alanine (top corner) with crystal growth.

The slower the solvent loss is from the droplets during the solvent exchange, the closer the pathway can be to the boundary.  At fast flow rates, S-L-L phase separation of the droplets became evident in cases when the entire droplets solidified.

\begin{figure}[htp]					\includegraphics[width=1.0\columnwidth]{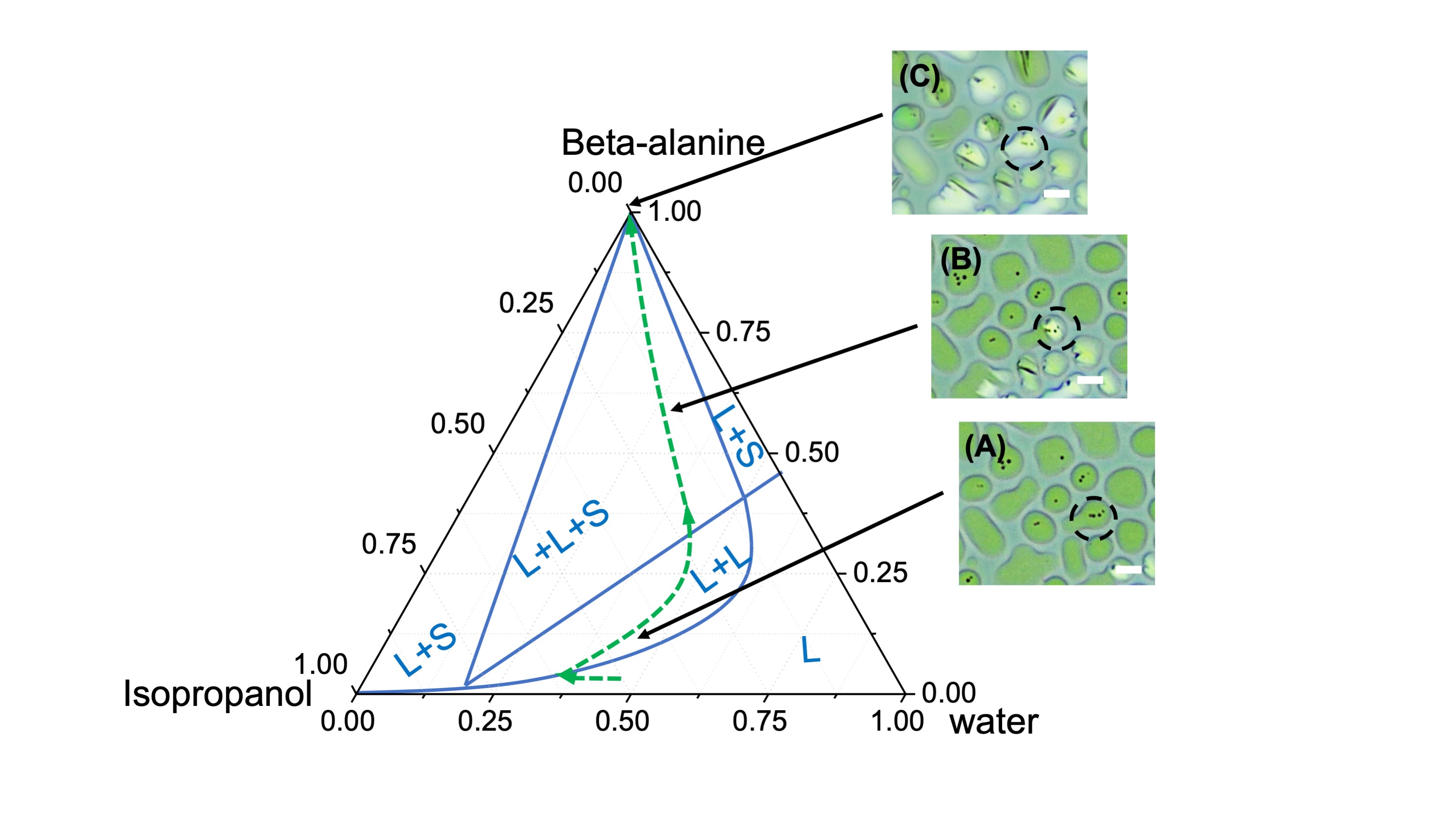}
				\caption{Crystallization in LLPS droplet inside the black dashed circle. (A) Before crystallization: tiny droplets inside big droplets (L-L region), (B) during the crystallization: the crystals inside the droplets (L-L-S region), (C) end of crystallization (only solid phase)} {with scale bar of 10 $\mu$m.} 
	\label{hole}
\end{figure}

Before crystallization, some small nanodroplets formed inside the droplet, for example, in Figure \ref{hole} at flow rate of 70 $ml/hr$. These tiny nanodroplets inside the droplets became defects (holes) inside the crystals. The hole formation suggest that the inner nanodroplets inside the droplet was alanine-lean phase that do not provide enough alanine for crystallization. These nanodroplets were different from crystal particles shown in Figure \ref{pathway} where the crystallization occurred at the edge of the droplets, and spread to the whole droplet.

 We can rationalize the dramatically different process in crystallization at different flow rates as following. The growth rate of crystals was determined by the rate for alanine to be incorporated in the existing crystals, and the rate for alanine to be transported to the boundary of the existing crystals. The relevant dimensionless number is Damkohler number (Da), the ratio between the reaction (here crystallization) rate and the rate of mass transport. If incorporation of alanine into crystals were the rate-limiting step for crystallization, the growth rate of crystals would not depend on the flow rate.  At faster flow rates, the growth rate of crystals was higher, suggesting that transport of alanine was the rate-limiting step. As the precursor alanine was confined in the droplets before crystallization, the rate-limiting step was the transport of good solvent (water) out from the droplets and of the antisolvent (isopropanol) into the droplets to reach the supersaturation for crystallization. The driving force for the flux into and out from the droplets was the chemical potential difference of solvents in droplets and in the flow. Higher flow rate accelerated the mass transfer from the droplets as the boundary layer thickness around the droplet ws compressed more by the external flow. The in-depth analysis for the similar scenario in droplet formation by solvent exchange has been reported in previous work \cite{zhang_2015_formation}. In other words, faster exchange rates have the same effect as faster cooling rate, creating a wider metastable zone width before crystallization.

The cascade crystallization of droplets in vicinity of growing crystals at high flow rates was attributed to the kinetics of nucleation at high {$S_{alanine}$} level. When a cluster of emulsion droplets hexadecane oil was cooled down in water with surfactant, the works of Abramov et al \cite{abramov_2016_crystallization} reported that at a high supercool rate, the droplets tended to crystallize in the neighborhood of already solidified droplets. In our experiments, the induced crystallization was attributed to local effect of the crystallization of neighbouring droplets. 
 
Surface wettability influences the formation and stability of droplets on the surface as well as the adhesion of the crystals on the surface. Water wets well the highly hydrophilic surface where the alanine-rich droplets do not preferentially nucleated, compared to the bulk nucleation. On the hydrophobic $OTS-Si$, the droplets were stable, but the adhesion of the crystals on the surface was low, so the crystals detached from the surface easily. 
 
Drastically varied shapes of crystals obtained at different flow rates demonstrate that the solvent exchange was an effective approach for controlling oiling-out crystallization on a substrate. The amount of droplet liquids determined how much crystals on the surface by solvent exchange.  At the same flow rate, there was good reproducibility in the main characteristics of the crystal shapes and surface coverage.  The produced crystals can be tuned to be thin film with a range of surface coverage at slow flow rates, or to be solid structures of different shapes at higher flow rates. The as-prepared crystals on surfaces may be used in a range of applications, such as seeds in crystallization, anti-fouling surfaces, or templates for functional structures.


\subsection{Potential application: surface crystals as seeds for crystallization in bulk}

{The interesting phenomena and different shapes of the crystals obtained in this research may have different applications, such as manufacturing film (2D) crystals or materials for special characteristics, screening the crystallization conditions for different crystal products. We note that one of the advantages for solvent exchange is to generate crystals with initially low concentration of alanine in the solution. For example, in this work with 1.5 \% alanine, it is easy to generate the crystals with solvent exchange, but it is nearly not possible to have crystals by adding anti-solvent. The crystals obtained can used as seeds to trigger crystallization.} In addition, it can be limited by availability of seed crystals, in particularly when an expensive new pharmaceutical compound can be only synthesized at a small quantity.  To facilitate purification of new compounds, it will be helpful to form crystals from a stock solution at a very low initial concentration. Here we show that oiling-out crystallization by solvent exchange can take place from a stock solution of low concentration and the crystals can be used as seeds for fast crystallization in bulk. 

\begin{figure} [htp]
	\includegraphics[width=1\columnwidth]{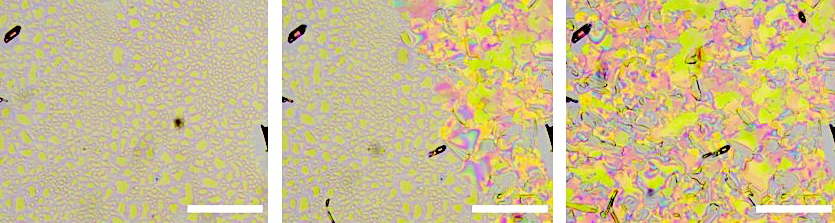}
	\caption{Optical microscopy of beta-alanine oiling out crystallization on {$SiO_2$} substrate after solvent exchange by using lower concentration of alanine(1.5 \%) in solution A with isopropanol flow rate of 12 {$ml/hr$}. Scale bar: 100 \(\mu m\)}	
	\label{lowC}
\end{figure}

 Oiling out crystallization was tested at 1.5 \% of alanine by weight in water and isopropanol mixture, much lower than the solutions of 3 \% for all the results above. Figure \ref{lowC} shows that lowering the initial concentration lead to thinner film crystals to form, due to combination of relatively small droplets formed during the liquid-liquid phase separation. {The alanine crystals were successfully generated by solvent exchange, but it would be hardly crystallized  by adding anti-solvent, because the crystallization pathway (connecting the initial starting point and the isopropanol point) would keep in homogeneous region (under-saturated) unless huge amount of isopropanol would be added.}

\begin{figure} [htp]
	\includegraphics[width=0.9\columnwidth]{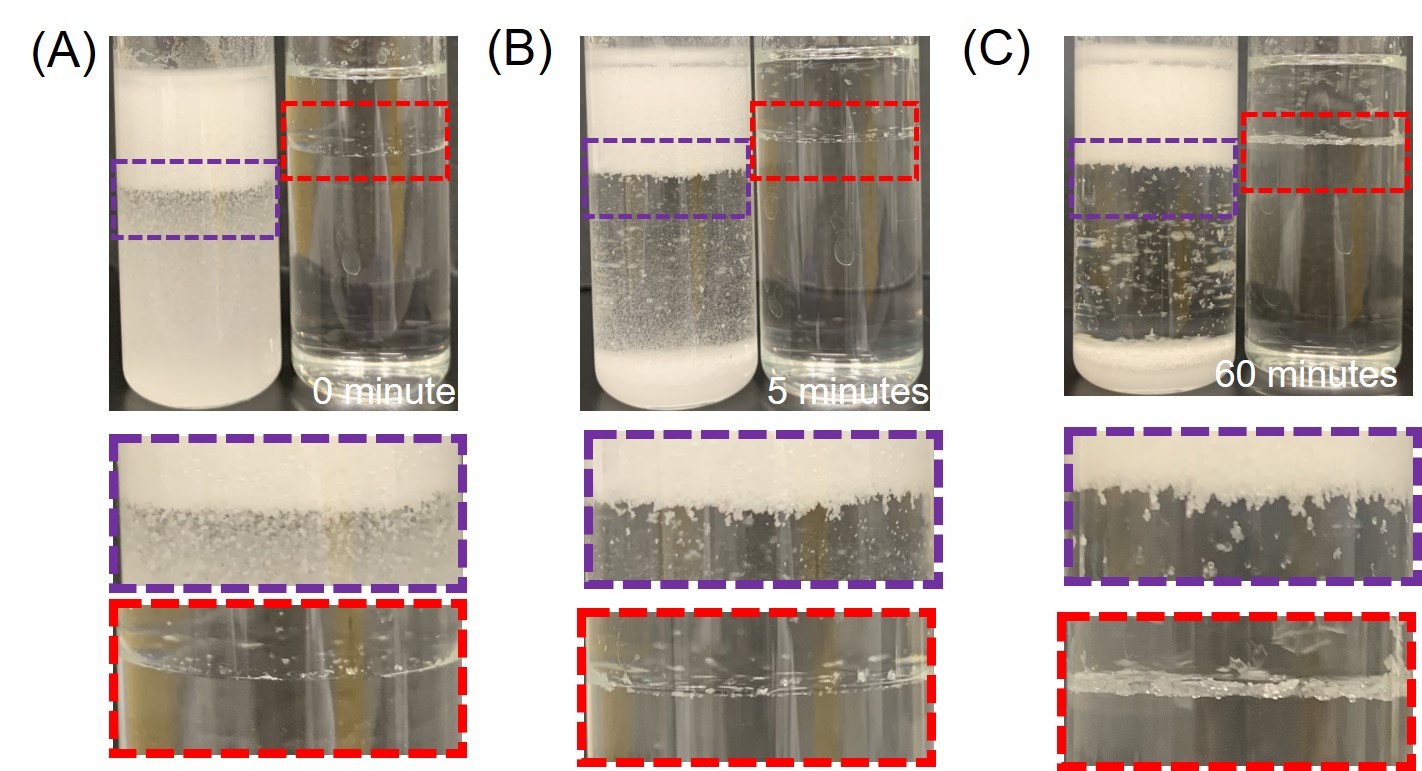}
	\caption{Crystallization in bulk solution with (left) and without (right) seeds. (A) immediately, (B) 5 minutes and (D) 60 minutes after mixing. The interface between two liquid phase is shown in the two images on the bottom.
}	
	\label{seed}
\end{figure}

 Immediately after mixing of the components at the mass ratio of 6.6 (alanine),8.8 (water) and 5.1 (isopropanol), only a small amount of crystals formed at the interface between two liquid phases as shown in Figure \ref{seed}. With time  more and more crystals formed slowly at the interface. After 1 hour, the amount of the crystals was still not enough to settle into the bottom phase. In contrast, the mixture with added crystal seeds at $5\times 10^{-5} g/ml$ formed a large amount of crystals immediately after mixing with both the top and bottom liquid layers turn to milky from the crystal particles. The results demonstrated that oiling-out crystallization by solvent exchange can produce sufficient amount of crystals that be used to trigger phase separation in bulk mixtures. 


{In this work, it was focused on the influence of flow rates. In the solvent exchange process, the change of the compositions in the droplets, determined by the flow rates, drove the nucleation and crystallization. The composition pathway in the ternary phase diagram was demonstrated for understanding the mechanism of oiling-out crystallization, i.e. crystallization in liquid-liquid-phase separation solution. The mechanism may be very complicated, the initial concentration of the alanine, the exchange solvent's compositions, together with the flow rates all will influence the formation of the droplets and the formation of the crystal products.   It is noted that the temperature is key thermodynamic factor for ternary phase diagram\cite{yang2014sandwich} \cite{du2016liquid},  it is also possible to design the crystallization pathways by cooling. It is important to further understand the LLPS crystallization by design different crystallization pathways in ternary diagram. With considering the influence of the surface properties \cite{yang2015phase}, the interactions between the solute-rich phase, solute-lean phase, flow phase and the surface properties could also influence the properties of the final products. These interactions and parameters influence the solvent exchange should be further investigated. In addition, the combination of other technologies such as Ultrasound \cite{gao2018ultrasonic} \cite{fang2020ultrasound} can also be useful to explore the oiling out / liquid-liquid phase separation crystallization, as well as its influences on the formation of droplets and the solvent exchange process. {In short, further studies in other parameters (such as surface micropattens, compositions of solutions and temperatures, affecting oiling-out crystallization, are still needed to advance understanding on oiling out crystallization by solvent exchange}.

\section{Conclusions}

 We demonstrated that the solvent exchange can be an unique approach for understanding oiling-out crystallization and producing crystals on solid surfaces. Primary features in morphology and surface coverage of crystals were highly reproducible in our experiments by adjusting the flow rate or wettability of the surface.  The morphology of the crystals can be thin films at slow flow rates, or droplet shapes at higher flow rates.  Particularly at fast flow rate, the crystals were constrained on the location of each droplet. {By solvent exchange, the solution with very low concentration of solute can be crystallised in a liquid-liquid phase separation system, which can not eaisly achieved by adding anti-solvent.} Importantly, {crystals obtained from} a solution with a low initial concentration of solute can be used as the seed to {accelerate the crystallization process} in a larger amount in bulk solution. 
 
  The promising results highlighted in study may advance understanding {the mechanisms on formation of both droplets and crystals in liquid-liquid phase separation system} and further investigations on these mechanisms} may inspire development of a new approach that may be valuable for controlling the processes in separation and purification of crystals or even surface functionalization for pharmaceutical and other applications.
 

\section{Acknowledge}
 The project is supported by the Natural Science and Engineering Research Council of Canada (NSERC) and Future Energy Systems (Canada First Research
Excellence Fund). This research was undertaken, in part, thanks to funding from the Canada Research Chairs program.

\bibliography{oilingout.bib}

\providecommand{\latin}[1]{#1}
\makeatletter
\providecommand{\doi}
  {\begingroup\let\do\@makeother\dospecials
  \catcode`\{=1 \catcode`\}=2 \doi@aux}
\providecommand{\doi@aux}[1]{\endgroup\texttt{#1}}
\makeatother
\providecommand*\mcitethebibliography{\thebibliography}
\csname @ifundefined\endcsname{endmcitethebibliography}
  {\let\endmcitethebibliography\endthebibliography}{}
\begin{mcitethebibliography}{57}
\providecommand*\natexlab[1]{#1}
\providecommand*\mciteSetBstSublistMode[1]{}
\providecommand*\mciteSetBstMaxWidthForm[2]{}
\providecommand*\mciteBstWouldAddEndPuncttrue
  {\def\EndOfBibitem{\unskip.}}
\providecommand*\mciteBstWouldAddEndPunctfalse
  {\let\EndOfBibitem\relax}
\providecommand*\mciteSetBstMidEndSepPunct[3]{}
\providecommand*\mciteSetBstSublistLabelBeginEnd[3]{}
\providecommand*\EndOfBibitem{}
\mciteSetBstSublistMode{f}
\mciteSetBstMaxWidthForm{subitem}{(\alph{mcitesubitemcount})}
\mciteSetBstSublistLabelBeginEnd
  {\mcitemaxwidthsubitemform\space}
  {\relax}
  {\relax}

\bibitem[Myerson(2002)]{myerson2002handbook}
Myerson,~A. \emph{Handbook of industrial crystallization};
  Butterworth-Heinemann, 2002\relax
\mciteBstWouldAddEndPuncttrue
\mciteSetBstMidEndSepPunct{\mcitedefaultmidpunct}
{\mcitedefaultendpunct}{\mcitedefaultseppunct}\relax
\EndOfBibitem
\bibitem[Kleinebudde \latin{et~al.}(2017)Kleinebudde, Khinast, and
  Rantanen]{kleinebudde2017continuous}
Kleinebudde,~P.; Khinast,~J.; Rantanen,~J. \emph{Continuous manufacturing of
  pharmaceuticals}; John Wiley \& Sons, 2017\relax
\mciteBstWouldAddEndPuncttrue
\mciteSetBstMidEndSepPunct{\mcitedefaultmidpunct}
{\mcitedefaultendpunct}{\mcitedefaultseppunct}\relax
\EndOfBibitem
\bibitem[Li \latin{et~al.}(2016)Li, Yin, Zhang, Hou, Bao, Gong, Hao, Wang,
  Wang, and Wang]{li_2016_process}
Li,~X.; Yin,~Q.; Zhang,~M.; Hou,~B.; Bao,~Y.; Gong,~J.; Hao,~H.; Wang,~Y.;
  Wang,~J.; Wang,~Z. Process design for antisolvent crystallization of
  erythromycin ethylsuccinate in oiling-out system. \emph{Industrial \&
  Engineering Chemistry Research} \textbf{2016}, \emph{55}, 7484--7492\relax
\mciteBstWouldAddEndPuncttrue
\mciteSetBstMidEndSepPunct{\mcitedefaultmidpunct}
{\mcitedefaultendpunct}{\mcitedefaultseppunct}\relax
\EndOfBibitem
\bibitem[Yu \latin{et~al.}(2005)Yu, Tan, and Chow]{yu_2005_effects}
Yu,~Z.; Tan,~R.; Chow,~P. Effects of operating conditions on agglomeration and
  habit of paracetamol crystals in anti-solvent crystallization. \emph{Journal
  of Crystal Growth} \textbf{2005}, \emph{279}, 477--488\relax
\mciteBstWouldAddEndPuncttrue
\mciteSetBstMidEndSepPunct{\mcitedefaultmidpunct}
{\mcitedefaultendpunct}{\mcitedefaultseppunct}\relax
\EndOfBibitem
\bibitem[Nowee \latin{et~al.}(2008)Nowee, Abbas, and
  Romagnoli]{nowee_2008_antisolvent}
Nowee,~S.~M.; Abbas,~A.; Romagnoli,~J.~A. Antisolvent crystallization: model
  identification, experimental validation and dynamic simulation.
  \emph{Chemical Engineering Science} \textbf{2008}, \emph{63},
  5457--5467\relax
\mciteBstWouldAddEndPuncttrue
\mciteSetBstMidEndSepPunct{\mcitedefaultmidpunct}
{\mcitedefaultendpunct}{\mcitedefaultseppunct}\relax
\EndOfBibitem
\bibitem[Kiesow \latin{et~al.}(2010)Kiesow, Ruether, and
  Sadowski]{kiesow_2010_solubility}
Kiesow,~K.; Ruether,~F.; Sadowski,~G. Solubility, crystallization and
  oiling-out behavior of PEGDME: 1. Pure-solvent systems. \emph{Fluid Phase
  Equilibria} \textbf{2010}, \emph{298}, 253--261\relax
\mciteBstWouldAddEndPuncttrue
\mciteSetBstMidEndSepPunct{\mcitedefaultmidpunct}
{\mcitedefaultendpunct}{\mcitedefaultseppunct}\relax
\EndOfBibitem
\bibitem[Lafferrère \latin{et~al.}(2004)Lafferrère, Hoff, and
  Veesler]{lafferrre_2004_study}
Lafferrère,~L.; Hoff,~C.; Veesler,~S. Study of liquid–liquid demixing from
  drug solution. \emph{Journal of Crystal Growth} \textbf{2004}, \emph{269},
  550--557\relax
\mciteBstWouldAddEndPuncttrue
\mciteSetBstMidEndSepPunct{\mcitedefaultmidpunct}
{\mcitedefaultendpunct}{\mcitedefaultseppunct}\relax
\EndOfBibitem
\bibitem[Suzuki \latin{et~al.}(2019)Suzuki, Sato, Nishiura, Harada, Kamasaka,
  Kuriki, and Tamura]{suzuki_2019_a}
Suzuki,~D.; Sato,~Y.; Nishiura,~H.; Harada,~R.; Kamasaka,~H.; Kuriki,~T.;
  Tamura,~H. A novel extraction method for aroma isolation from dark chocolate
  based on the oiling-out effect. \emph{Food Analytical Methods} \textbf{2019},
  \emph{12}, 2857--2869\relax
\mciteBstWouldAddEndPuncttrue
\mciteSetBstMidEndSepPunct{\mcitedefaultmidpunct}
{\mcitedefaultendpunct}{\mcitedefaultseppunct}\relax
\EndOfBibitem
\bibitem[Bhamidi and Abolins(2019)Bhamidi, and Abolins]{bhamidi_2019_a}
Bhamidi,~V.; Abolins,~B.~P. A thermodynamic approach for the prediction of
  oiling out boundaries from solubility data. \emph{Processes} \textbf{2019},
  \emph{7}, 577\relax
\mciteBstWouldAddEndPuncttrue
\mciteSetBstMidEndSepPunct{\mcitedefaultmidpunct}
{\mcitedefaultendpunct}{\mcitedefaultseppunct}\relax
\EndOfBibitem
\bibitem[Yang and Rasmuson(2014)Yang, and Rasmuson]{yang2014ternary}
Yang,~H.; Rasmuson,~{\AA}.~C. Ternary phase diagrams of ethyl paraben and
  propyl paraben in ethanol aqueous solvents. \emph{Fluid Phase Equilibria}
  \textbf{2014}, \emph{376}, 69--75\relax
\mciteBstWouldAddEndPuncttrue
\mciteSetBstMidEndSepPunct{\mcitedefaultmidpunct}
{\mcitedefaultendpunct}{\mcitedefaultseppunct}\relax
\EndOfBibitem
\bibitem[Yang and Rasmuson(2012)Yang, and Rasmuson]{yang2012investigation}
Yang,~H.; Rasmuson,~{\AA}.~C. Investigation of batch cooling crystallization in
  a liquid--liquid separating system by PAT. \emph{Organic Process Research \&
  Development} \textbf{2012}, \emph{16}, 1212--1224\relax
\mciteBstWouldAddEndPuncttrue
\mciteSetBstMidEndSepPunct{\mcitedefaultmidpunct}
{\mcitedefaultendpunct}{\mcitedefaultseppunct}\relax
\EndOfBibitem
\bibitem[Ilevbare and Taylor(2013)Ilevbare, and
  Taylor]{ilevbare_2013_liquidliquid}
Ilevbare,~G.~A.; Taylor,~L.~S. Liquid–liquid phase separation in highly
  supersaturated aqueous solutions of poorly water-soluble drugs: implications
  for solubility enhancing formulations. \emph{Crystal Growth \& Design}
  \textbf{2013}, \emph{13}, 1497--1509\relax
\mciteBstWouldAddEndPuncttrue
\mciteSetBstMidEndSepPunct{\mcitedefaultmidpunct}
{\mcitedefaultendpunct}{\mcitedefaultseppunct}\relax
\EndOfBibitem
\bibitem[Serajudin and Pudipeddi(2002)Serajudin, and
  Pudipeddi]{serajudin_2002_salt}
Serajudin,~A. T.~M.; Pudipeddi,~M. \emph{Salt selection strategies, in Handbook
  of Pharmaceutical Salts: Properties, Selection And Use}, stahl, p.h. and
  wermuth, c.g., wermuth, c.g. ed.; VHCA, Verlag Helvetica Chimica
  Acta/Wiley–VCH, Zürich/Weinheim, 2002; pp 135 -- 160\relax
\mciteBstWouldAddEndPuncttrue
\mciteSetBstMidEndSepPunct{\mcitedefaultmidpunct}
{\mcitedefaultendpunct}{\mcitedefaultseppunct}\relax
\EndOfBibitem
\bibitem[de~Albuquerque and Mazzotti(2017)de~Albuquerque, and
  Mazzotti]{dealbuquerque_2017_influence}
de~Albuquerque,~I.; Mazzotti,~M. Influence of liquid-liquid phase separation on
  the crystallization of l -menthol from water. \emph{Chemical Engineering \&
  Technology} \textbf{2017}, \emph{40}, 1339--1346\relax
\mciteBstWouldAddEndPuncttrue
\mciteSetBstMidEndSepPunct{\mcitedefaultmidpunct}
{\mcitedefaultendpunct}{\mcitedefaultseppunct}\relax
\EndOfBibitem
\bibitem[Sun \latin{et~al.}(2020)Sun, Du, Yang, Wang, Gao, and
  Gong]{sun2020understanding}
Sun,~M.; Du,~S.; Yang,~J.; Wang,~L.; Gao,~Z.; Gong,~J. Understanding the
  effects of upstream impurities on the oiling-out and crystallization of
  $\gamma$-aminobutyric acid. \emph{Organic Process Research \& Development}
  \textbf{2020}, \emph{24}, 398--404\relax
\mciteBstWouldAddEndPuncttrue
\mciteSetBstMidEndSepPunct{\mcitedefaultmidpunct}
{\mcitedefaultendpunct}{\mcitedefaultseppunct}\relax
\EndOfBibitem
\bibitem[Tanaka and Takiyama(2019)Tanaka, and Takiyama]{tanaka2019effect}
Tanaka,~K.; Takiyama,~H. Effect of oiling-out during crystallization on
  purification of an intermediate compound. \emph{Organic Process Research \&
  Development} \textbf{2019}, \emph{23}, 2001--2008\relax
\mciteBstWouldAddEndPuncttrue
\mciteSetBstMidEndSepPunct{\mcitedefaultmidpunct}
{\mcitedefaultendpunct}{\mcitedefaultseppunct}\relax
\EndOfBibitem
\bibitem[Wang \latin{et~al.}(2020)Wang, Cao, Zhu, Wang, and
  Lakerveld]{wang2020emulsion}
Wang,~J.; Cao,~W.; Zhu,~L.; Wang,~J.; Lakerveld,~R. Emulsion-assisted cooling
  crystallization of ibuprofen. \emph{Chemical Engineering Science}
  \textbf{2020}, 115861\relax
\mciteBstWouldAddEndPuncttrue
\mciteSetBstMidEndSepPunct{\mcitedefaultmidpunct}
{\mcitedefaultendpunct}{\mcitedefaultseppunct}\relax
\EndOfBibitem
\bibitem[Bonnett \latin{et~al.}(2003)Bonnett, Carpenter, Dawson, and
  Davey]{bonnett_2003_solution}
Bonnett,~P.~E.; Carpenter,~K.~J.; Dawson,~S.; Davey,~R.~J. Solution
  crystallisation via a submerged liquid–liquid phase boundary: oiling out.
  \emph{Chemical Communications} \textbf{2003}, 698--699\relax
\mciteBstWouldAddEndPuncttrue
\mciteSetBstMidEndSepPunct{\mcitedefaultmidpunct}
{\mcitedefaultendpunct}{\mcitedefaultseppunct}\relax
\EndOfBibitem
\bibitem[Sun \latin{et~al.}(2018)Sun, Du, Chen, Rohani, Zhang, Liu, Sun, Wang,
  Shi, Xu, and Gong]{sun_2018_oilingout}
Sun,~M.; Du,~S.; Chen,~M.; Rohani,~S.; Zhang,~H.; Liu,~Y.; Sun,~P.; Wang,~Y.;
  Shi,~P.; Xu,~S.; Gong,~J. Oiling-out investigation and morphology control of
  $\beta$-alanine based on ternary phase diagrams. \emph{Crystal Growth \&
  Design} \textbf{2018}, \emph{18}, 818--826\relax
\mciteBstWouldAddEndPuncttrue
\mciteSetBstMidEndSepPunct{\mcitedefaultmidpunct}
{\mcitedefaultendpunct}{\mcitedefaultseppunct}\relax
\EndOfBibitem
\bibitem[Sun \latin{et~al.}(2018)Sun, Tang, Du, Zhang, Fu, and
  Gong]{sun_2018_understanding}
Sun,~M.; Tang,~W.; Du,~S.; Zhang,~Y.; Fu,~X.; Gong,~J. Understanding the roles
  of oiling-out on crystallization of $\beta$-alanine: unusual behavior in
  metastable zone width and surface nucleation during growth stage.
  \emph{Crystal Growth \& Design} \textbf{2018}, \emph{18}, 6885--6890\relax
\mciteBstWouldAddEndPuncttrue
\mciteSetBstMidEndSepPunct{\mcitedefaultmidpunct}
{\mcitedefaultendpunct}{\mcitedefaultseppunct}\relax
\EndOfBibitem
\bibitem[Pitt \latin{et~al.}(2018)Pitt, Peña, Tew, Pal, Smith, Nagy, and
  Litster]{pitt_2018_particle}
Pitt,~K.; Peña,~R.; Tew,~J.~D.; Pal,~K.; Smith,~R.; Nagy,~Z.~K.;
  Litster,~J.~D. Particle design via spherical agglomeration: A critical review
  of controlling parameters, rate processes and modelling. \emph{Powder
  Technology} \textbf{2018}, \emph{326}, 327--343\relax
\mciteBstWouldAddEndPuncttrue
\mciteSetBstMidEndSepPunct{\mcitedefaultmidpunct}
{\mcitedefaultendpunct}{\mcitedefaultseppunct}\relax
\EndOfBibitem
\bibitem[Kawashima \latin{et~al.}(1982)Kawashima, Okumura, and
  Takenaka]{kawashima_1982_spherical}
Kawashima,~Y.; Okumura,~M.; Takenaka,~H. Spherical crystallization: direct
  spherical agglomeration of salicylic acid crystals during crystallization.
  \emph{Science} \textbf{1982}, \emph{216}, 1127--1128\relax
\mciteBstWouldAddEndPuncttrue
\mciteSetBstMidEndSepPunct{\mcitedefaultmidpunct}
{\mcitedefaultendpunct}{\mcitedefaultseppunct}\relax
\EndOfBibitem
\bibitem[Maeda \latin{et~al.}(1997)Maeda, Nomura, Fukui, and
  Hirota]{maeda_1997_separation}
Maeda,~K.; Nomura,~Y.; Fukui,~K.; Hirota,~S. Separation of fatty acids by
  crystallization using two liquid phases. \emph{Korean Journal of Chemical
  Engineering} \textbf{1997}, \emph{14}, 175--178\relax
\mciteBstWouldAddEndPuncttrue
\mciteSetBstMidEndSepPunct{\mcitedefaultmidpunct}
{\mcitedefaultendpunct}{\mcitedefaultseppunct}\relax
\EndOfBibitem
\bibitem[Sun \latin{et~al.}(2019)Sun, Shichao, Tang, Jia, and
  Gong]{sun2019design}
Sun,~M.; Shichao,~D.; Tang,~W.; Jia,~L.; Gong,~J. Design of spherical
  crystallization for drugs based on thermal-induced liquid--liquid phase
  separation: case studies of water-insoluble drugs. \emph{Industrial \&
  Engineering Chemistry Research} \textbf{2019}, \emph{58}, 20401--20411\relax
\mciteBstWouldAddEndPuncttrue
\mciteSetBstMidEndSepPunct{\mcitedefaultmidpunct}
{\mcitedefaultendpunct}{\mcitedefaultseppunct}\relax
\EndOfBibitem
\bibitem[Duffy \latin{et~al.}(2012)Duffy, Cremin, Napier, Robinson, Barrett,
  Hao, and Glennon]{duffy2012situ}
Duffy,~D.; Cremin,~N.; Napier,~M.; Robinson,~S.; Barrett,~M.; Hao,~H.;
  Glennon,~B. In situ monitoring, control and optimization of a liquid--liquid
  phase separation crystallization. \emph{Chemical engineering science}
  \textbf{2012}, \emph{77}, 112--121\relax
\mciteBstWouldAddEndPuncttrue
\mciteSetBstMidEndSepPunct{\mcitedefaultmidpunct}
{\mcitedefaultendpunct}{\mcitedefaultseppunct}\relax
\EndOfBibitem
\bibitem[Yang \latin{et~al.}(2014)Yang, Chen, and Rasmuson]{yang2014sandwich}
Yang,~H.; Chen,~H.; Rasmuson,~{\AA}.~C. Sandwich crystals of butyl paraben.
  \emph{CrystEngComm} \textbf{2014}, \emph{16}, 8863--8873\relax
\mciteBstWouldAddEndPuncttrue
\mciteSetBstMidEndSepPunct{\mcitedefaultmidpunct}
{\mcitedefaultendpunct}{\mcitedefaultseppunct}\relax
\EndOfBibitem
\bibitem[Deneau and Steele(2005)Deneau, and Steele]{deneau_2005_an}
Deneau,~E.; Steele,~G. An in-line study of oiling out and crystallization.
  \emph{Organic Process Research \& Development} \textbf{2005}, \emph{9},
  943--950\relax
\mciteBstWouldAddEndPuncttrue
\mciteSetBstMidEndSepPunct{\mcitedefaultmidpunct}
{\mcitedefaultendpunct}{\mcitedefaultseppunct}\relax
\EndOfBibitem
\bibitem[McClements(2012)]{mcclements_2012_crystals}
McClements,~D.~J. Crystals and crystallization in oil-in-water emulsions:
  Implications for emulsion-based delivery systems. \emph{Advances in Colloid
  and Interface Science} \textbf{2012}, \emph{174}, 1--30\relax
\mciteBstWouldAddEndPuncttrue
\mciteSetBstMidEndSepPunct{\mcitedefaultmidpunct}
{\mcitedefaultendpunct}{\mcitedefaultseppunct}\relax
\EndOfBibitem
\bibitem[Codan \latin{et~al.}(2010)Codan, Bäbler, and
  Mazzotti]{codan2010phase}
Codan,~L.; Bäbler,~M.~U.; Mazzotti,~M. Phase diagram of a chiral substance
  exhibiting oiling out in cyclohexane. \emph{Crystal growth \& design}
  \textbf{2010}, \emph{10}, 4005--4013\relax
\mciteBstWouldAddEndPuncttrue
\mciteSetBstMidEndSepPunct{\mcitedefaultmidpunct}
{\mcitedefaultendpunct}{\mcitedefaultseppunct}\relax
\EndOfBibitem
\bibitem[Goetsch \latin{et~al.}(2016)Goetsch, Zimmermann, Van Den~Bongard,
  Enders, and Zeiner]{goetsch2016superposition}
Goetsch,~T.; Zimmermann,~P.; Van Den~Bongard,~R.; Enders,~S.; Zeiner,~T.
  Superposition of liquid--liquid and solid--liquid equilibria of linear and
  branched molecules: binary systems. \emph{Industrial \& engineering chemistry
  research} \textbf{2016}, \emph{55}, 11167--11174\relax
\mciteBstWouldAddEndPuncttrue
\mciteSetBstMidEndSepPunct{\mcitedefaultmidpunct}
{\mcitedefaultendpunct}{\mcitedefaultseppunct}\relax
\EndOfBibitem
\bibitem[Tatsukawa \latin{et~al.}(2020)Tatsukawa, Kadota, Yoshida, and
  Shirakawa]{tatsukawa2020development}
Tatsukawa,~S.; Kadota,~K.; Yoshida,~M.; Shirakawa,~Y. Development of
  quantifying supersaturation to determine the effect of the anti-solvent on
  precipitation in liquid-liquid interfacial crystallization. \emph{Journal of
  Molecular Liquids} \textbf{2020}, 113097\relax
\mciteBstWouldAddEndPuncttrue
\mciteSetBstMidEndSepPunct{\mcitedefaultmidpunct}
{\mcitedefaultendpunct}{\mcitedefaultseppunct}\relax
\EndOfBibitem
\bibitem[Ianiro \latin{et~al.}(2019)Ianiro, Wu, van Rijt, Vena, Keizer,
  Esteves, Tuinier, Friedrich, Sommerdijk, and Patterson]{ianiro2019liquid}
Ianiro,~A.; Wu,~H.; van Rijt,~M.~M.; Vena,~M.~P.; Keizer,~A.~D.; Esteves,~A.
  C.~C.; Tuinier,~R.; Friedrich,~H.; Sommerdijk,~N.~A.; Patterson,~J.~P.
  Liquid--liquid phase separation during amphiphilic self-assembly.
  \emph{Nature chemistry} \textbf{2019}, \emph{11}, 320--328\relax
\mciteBstWouldAddEndPuncttrue
\mciteSetBstMidEndSepPunct{\mcitedefaultmidpunct}
{\mcitedefaultendpunct}{\mcitedefaultseppunct}\relax
\EndOfBibitem
\bibitem[Walton and Wynne(2019)Walton, and Wynne]{walton2019using}
Walton,~F.; Wynne,~K. Using optical tweezing to control phase separation and
  nucleation near a liquid--liquid critical point. \emph{Soft Matter}
  \textbf{2019}, \emph{15}, 8279--8289\relax
\mciteBstWouldAddEndPuncttrue
\mciteSetBstMidEndSepPunct{\mcitedefaultmidpunct}
{\mcitedefaultendpunct}{\mcitedefaultseppunct}\relax
\EndOfBibitem
\bibitem[Zhang and Ducker(2007)Zhang, and Ducker]{zhang2007}
Zhang,~X.~H.; Ducker,~W. Formation of interfacial nanodroplets through changes
  in solvent quality. \emph{Langmuir} \textbf{2007}, \emph{23},
  12478--12480\relax
\mciteBstWouldAddEndPuncttrue
\mciteSetBstMidEndSepPunct{\mcitedefaultmidpunct}
{\mcitedefaultendpunct}{\mcitedefaultseppunct}\relax
\EndOfBibitem
\bibitem[Zhang \latin{et~al.}(2015)Zhang, Lu, Tan, Bao, He, Sun, and
  Lohse]{zhang_2015_formation}
Zhang,~X.; Lu,~Z.; Tan,~H.; Bao,~L.; He,~Y.; Sun,~C.; Lohse,~D. Formation of
  surface nanodroplets under controlled flow conditions. \emph{Proceedings of
  the National Academy of Sciences} \textbf{2015}, \emph{112}, 9253--9257\relax
\mciteBstWouldAddEndPuncttrue
\mciteSetBstMidEndSepPunct{\mcitedefaultmidpunct}
{\mcitedefaultendpunct}{\mcitedefaultseppunct}\relax
\EndOfBibitem
\bibitem[Bao \latin{et~al.}(2016)Bao, Werbiuk, Lohse, and Zhang]{bao2016}
Bao,~L.; Werbiuk,~Z.; Lohse,~D.; Zhang,~X. Controlling the growth modes of
  femtoliter sessile droplets nucleating on chemically patterned surfaces.
  \emph{The Journal of Physical Chemistry Letters} \textbf{2016}, \emph{7},
  1055--1059\relax
\mciteBstWouldAddEndPuncttrue
\mciteSetBstMidEndSepPunct{\mcitedefaultmidpunct}
{\mcitedefaultendpunct}{\mcitedefaultseppunct}\relax
\EndOfBibitem
\bibitem[Li \latin{et~al.}(2019)Li, Dyett, and Zhang]{li2019}
Li,~M.; Dyett,~B.; Zhang,~X. Automated femtoliter droplet-based determination
  of oil water partition coefficient. \emph{Analytical Chemistry}
  \textbf{2019}, \emph{91}, 10371--10375\relax
\mciteBstWouldAddEndPuncttrue
\mciteSetBstMidEndSepPunct{\mcitedefaultmidpunct}
{\mcitedefaultendpunct}{\mcitedefaultseppunct}\relax
\EndOfBibitem
\bibitem[Li \latin{et~al.}(2019)Li, Dyett, Yu, Bansal, and Zhang]{li2019small}
Li,~M.; Dyett,~B.; Yu,~H.; Bansal,~V.; Zhang,~X. Functional femtoliter droplets
  for ultrafast nanoextraction and supersensitive online microanalysis.
  \emph{Small} \textbf{2019}, \emph{15}, 1804683\relax
\mciteBstWouldAddEndPuncttrue
\mciteSetBstMidEndSepPunct{\mcitedefaultmidpunct}
{\mcitedefaultendpunct}{\mcitedefaultseppunct}\relax
\EndOfBibitem
\bibitem[Dyett \latin{et~al.}(2018)Dyett, Kiyama, Rump, Tagawa, Lohse, and
  Zhang]{dyett2018}
Dyett,~B.; Kiyama,~A.; Rump,~M.; Tagawa,~Y.; Lohse,~D.; Zhang,~X. Growth
  dynamics of surface nanodroplets during solvent exchange at varying flow
  rates. \emph{Soft Matter} \textbf{2018}, \emph{14}, 5197--5204\relax
\mciteBstWouldAddEndPuncttrue
\mciteSetBstMidEndSepPunct{\mcitedefaultmidpunct}
{\mcitedefaultendpunct}{\mcitedefaultseppunct}\relax
\EndOfBibitem
\bibitem[Bao \latin{et~al.}(2015)Bao, Rezk, Yeo, and Zhang]{bao2015}
Bao,~L.; Rezk,~A.~R.; Yeo,~L.~Y.; Zhang,~X. Highly ordered arrays of femtoliter
  surface droplets. \emph{Small} \textbf{2015}, \emph{11}, 4850--4855\relax
\mciteBstWouldAddEndPuncttrue
\mciteSetBstMidEndSepPunct{\mcitedefaultmidpunct}
{\mcitedefaultendpunct}{\mcitedefaultseppunct}\relax
\EndOfBibitem
\bibitem[Yang and Rasmuson(2015)Yang, and Rasmuson]{yang2015phase}
Yang,~H.; Rasmuson,~{\AA}.~C. Phase equilibrium and mechanisms of
  crystallization in liquid--liquid phase separating system. \emph{Fluid Phase
  Equilibria} \textbf{2015}, \emph{385}, 120--128\relax
\mciteBstWouldAddEndPuncttrue
\mciteSetBstMidEndSepPunct{\mcitedefaultmidpunct}
{\mcitedefaultendpunct}{\mcitedefaultseppunct}\relax
\EndOfBibitem
\bibitem[Li \latin{et~al.}(0)Li, Salvator, Wijshoff, Versluis, and
  Lohse]{li2020}
Li,~Y.; Salvator,~V.; Wijshoff,~H.; Versluis,~M.; Lohse,~D. Evaporation-induced
  crystallization of surfactants in sessile multicomponent droplets.
  \emph{Langmuir} \textbf{0}, \emph{0}, null\relax
\mciteBstWouldAddEndPuncttrue
\mciteSetBstMidEndSepPunct{\mcitedefaultmidpunct}
{\mcitedefaultendpunct}{\mcitedefaultseppunct}\relax
\EndOfBibitem
\bibitem[Shanthi \latin{et~al.}(2013)Shanthi, Selvarajan, HemaDurga, and Lincy
  Mary~Ponmani]{shanthi_2013_nucleation}
Shanthi,~D.; Selvarajan,~P.; HemaDurga,~K.; Lincy Mary~Ponmani,~S. Nucleation
  kinetics, growth and studies of $\beta$-alanine single crystals.
  \emph{Spectrochimica Acta Part A: Molecular and Biomolecular Spectroscopy}
  \textbf{2013}, \emph{110}, 1--6\relax
\mciteBstWouldAddEndPuncttrue
\mciteSetBstMidEndSepPunct{\mcitedefaultmidpunct}
{\mcitedefaultendpunct}{\mcitedefaultseppunct}\relax
\EndOfBibitem
\bibitem[Zhang \latin{et~al.}(2008)Zhang, Quinn, and Ducker]{zhang2008}
Zhang,~X.~H.; Quinn,~A.; Ducker,~W.~A. Nanobubbles at the interface between
  water and a hydrophobic solid. \emph{Langmuir} \textbf{2008}, \emph{24},
  4756--4764\relax
\mciteBstWouldAddEndPuncttrue
\mciteSetBstMidEndSepPunct{\mcitedefaultmidpunct}
{\mcitedefaultendpunct}{\mcitedefaultseppunct}\relax
\EndOfBibitem
\bibitem[Dyett \latin{et~al.}(2017)Dyett, Yu, and Zhang]{fluidchamber}
Dyett,~B.; Yu,~H.; Zhang,~X. Formation of surface nanodroplets of viscous
  liquids by solvent exchange. \emph{The European Physical Journal E}
  \textbf{2017}, \emph{40}, 1--6\relax
\mciteBstWouldAddEndPuncttrue
\mciteSetBstMidEndSepPunct{\mcitedefaultmidpunct}
{\mcitedefaultendpunct}{\mcitedefaultseppunct}\relax
\EndOfBibitem
\bibitem[Zhang \latin{et~al.}(2015)Zhang, Wang, Bao, Dietrich, van~der Veen,
  Peng, Friend, Zandvliet, Yeo, and Lohse]{zhang_2015_mixed}
Zhang,~X.; Wang,~J.; Bao,~L.; Dietrich,~E.; van~der Veen,~R. C.~A.; Peng,~S.;
  Friend,~J.; Zandvliet,~H. J.~W.; Yeo,~L.; Lohse,~D. Mixed mode of dissolving
  immersed nanodroplets at a solid–water interface. \emph{Soft Matter}
  \textbf{2015}, \emph{11}, 1889--1900\relax
\mciteBstWouldAddEndPuncttrue
\mciteSetBstMidEndSepPunct{\mcitedefaultmidpunct}
{\mcitedefaultendpunct}{\mcitedefaultseppunct}\relax
\EndOfBibitem
\bibitem[Paul \latin{et~al.}(2020)Paul, Taylor, Murphy, Krzyzaniak, Dawson,
  Mullarney, Meenan, and Sun]{paul2020toward}
Paul,~S.; Taylor,~L.~J.; Murphy,~B.; Krzyzaniak,~J.~F.; Dawson,~N.;
  Mullarney,~M.~P.; Meenan,~P.; Sun,~C.~C. Toward a Molecular Understanding of
  the Impact of Crystal Size and Shape on Punch Sticking. \emph{Molecular
  Pharmaceutics} \textbf{2020}, \emph{17}, 1148--1158\relax
\mciteBstWouldAddEndPuncttrue
\mciteSetBstMidEndSepPunct{\mcitedefaultmidpunct}
{\mcitedefaultendpunct}{\mcitedefaultseppunct}\relax
\EndOfBibitem
\bibitem[Guzzo \latin{et~al.}(2019)Guzzo, de~Barros, and
  de~Arruda~Tino]{guzzo2019effect}
Guzzo,~P.~L.; de~Barros,~F. B.~M.; de~Arruda~Tino,~A.~A. Effect of prolonged
  dry grinding on size distribution, crystal structure and thermal
  decomposition of ultrafine particles of dolostone. \emph{Powder Technology}
  \textbf{2019}, \emph{342}, 141--148\relax
\mciteBstWouldAddEndPuncttrue
\mciteSetBstMidEndSepPunct{\mcitedefaultmidpunct}
{\mcitedefaultendpunct}{\mcitedefaultseppunct}\relax
\EndOfBibitem
\bibitem[Lu \latin{et~al.}(2016)Lu, Peng, and Zhang]{lu2016}
Lu,~Z.; Peng,~S.; Zhang,~X. Influence of solution composition on the formation
  of surface nanodroplets by solvent exchange. \emph{Langmuir} \textbf{2016},
  \emph{32}, 1700--1706\relax
\mciteBstWouldAddEndPuncttrue
\mciteSetBstMidEndSepPunct{\mcitedefaultmidpunct}
{\mcitedefaultendpunct}{\mcitedefaultseppunct}\relax
\EndOfBibitem
\bibitem[Xu \latin{et~al.}(2017)Xu, Yu, Peng, Lu, Lei, Lohse, and
  Zhang]{xue2017}
Xu,~C.; Yu,~H.; Peng,~S.; Lu,~Z.; Lei,~L.; Lohse,~D.; Zhang,~X. Collective
  interactions in the nucleation and growth of surface droplets. \emph{Soft
  Matter} \textbf{2017}, \emph{13}, 937--944\relax
\mciteBstWouldAddEndPuncttrue
\mciteSetBstMidEndSepPunct{\mcitedefaultmidpunct}
{\mcitedefaultendpunct}{\mcitedefaultseppunct}\relax
\EndOfBibitem
\bibitem[Mullin(2012)]{mullin2012industrial}
Mullin,~J. \emph{Industrial crystallization}; Springer Science \& Business
  Media, 2012\relax
\mciteBstWouldAddEndPuncttrue
\mciteSetBstMidEndSepPunct{\mcitedefaultmidpunct}
{\mcitedefaultendpunct}{\mcitedefaultseppunct}\relax
\EndOfBibitem
\bibitem[Kant \latin{et~al.}(2020)Kant, Koldeweij, Harth, van Limbeek, and
  Lohse]{Kant2020}
Kant,~P.; Koldeweij,~R. B.~J.; Harth,~K.; van Limbeek,~M. A.~J.; Lohse,~D.
  Fast-freezing kinetics inside a droplet impacting on a cold surface.
  \emph{Proceedings of the National Academy of Sciences} \textbf{2020},
  \emph{117}, 2788--2794\relax
\mciteBstWouldAddEndPuncttrue
\mciteSetBstMidEndSepPunct{\mcitedefaultmidpunct}
{\mcitedefaultendpunct}{\mcitedefaultseppunct}\relax
\EndOfBibitem
\bibitem[Abramov \latin{et~al.}(2016)Abramov, Ruppik, and
  Schuchmann]{abramov_2016_crystallization}
Abramov,~S.; Ruppik,~P.; Schuchmann,~H. Crystallization in emulsions: a
  thermo-optical method to determine single crystallization events in droplet
  clusters. \emph{Processes} \textbf{2016}, \emph{4}, 25\relax
\mciteBstWouldAddEndPuncttrue
\mciteSetBstMidEndSepPunct{\mcitedefaultmidpunct}
{\mcitedefaultendpunct}{\mcitedefaultseppunct}\relax
\EndOfBibitem
\bibitem[Du \latin{et~al.}(2016)Du, Wang, Du, Wang, Huang, Qin, and
  Gong]{du2016liquid}
Du,~Y.; Wang,~H.; Du,~S.; Wang,~Y.; Huang,~C.; Qin,~Y.; Gong,~J. The
  liquid--liquid phase separation and crystallization of vanillin in
  1-propanol/water solution. \emph{Fluid Phase Equilibria} \textbf{2016},
  \emph{409}, 84--91\relax
\mciteBstWouldAddEndPuncttrue
\mciteSetBstMidEndSepPunct{\mcitedefaultmidpunct}
{\mcitedefaultendpunct}{\mcitedefaultseppunct}\relax
\EndOfBibitem
\bibitem[Gao \latin{et~al.}(2018)Gao, Altimimi, Gong, Bao, Wang, and
  Rohani]{gao2018ultrasonic}
Gao,~Z.; Altimimi,~F.; Gong,~J.; Bao,~Y.; Wang,~J.; Rohani,~S. Ultrasonic
  irradiation and seeding to prevent metastable liquid--liquid phase separation
  and intensify crystallization. \emph{Crystal Growth \& Design} \textbf{2018},
  \emph{18}, 2628--2635\relax
\mciteBstWouldAddEndPuncttrue
\mciteSetBstMidEndSepPunct{\mcitedefaultmidpunct}
{\mcitedefaultendpunct}{\mcitedefaultseppunct}\relax
\EndOfBibitem
\bibitem[Fang \latin{et~al.}(2020)Fang, Tang, Wu, Wang, Gao, and
  Gong]{fang2020ultrasound}
Fang,~C.; Tang,~W.; Wu,~S.; Wang,~J.; Gao,~Z.; Gong,~J. Ultrasound-assisted
  intensified crystallization of L-glutamic acid: Crystal nucleation and
  polymorph transformation. \emph{Ultrasonics Sonochemistry} \textbf{2020},
  \emph{68}, 105227\relax
\mciteBstWouldAddEndPuncttrue
\mciteSetBstMidEndSepPunct{\mcitedefaultmidpunct}
{\mcitedefaultendpunct}{\mcitedefaultseppunct}\relax
\EndOfBibitem
\end{mcitethebibliography}

\end{document}